%% file: iclr2027_conference.tex
\documentclass{article} 
\usepackage{iclr2027_conference,times}

\input{math_commands.tex}

\usepackage{hyperref}
\usepackage{url}
\usepackage{float}
\usepackage[utf8]{inputenc} 
\usepackage[T1]{fontenc}    
\usepackage{hyperref}       
\usepackage{url}            
\usepackage{booktabs}       
\usepackage{amsfonts}       
\usepackage{nicefrac}       
\usepackage{microtype}      
\usepackage{xcolor}         
\usepackage{booktabs}
\usepackage{graphicx}
\usepackage{amsfonts}
\usepackage{amsmath}
\usepackage{amssymb}
\usepackage{multirow}
\usepackage{comment}
\usepackage{wrapfig}
\title{GLAD: Global-Local Adaptive Detector for Robust Speech Deepfake Detection}

\author{
Zelin Zhao\thanks{Equal contribution.} \quad
Guanjie Huang\footnotemark[1] \quad
Danny Hin Kwok Tsang \quad
Li Liu\thanks{Corresponding author: avrillliu@hkust-gz.edu.cn} \\
The Hong Kong University of Science and Technology (Guangzhou) \\
Guangzhou, China \\
\texttt{\{zzhao450,ghuang565\}@connect.hkust-gz.edu.cn} \\
\texttt{\{eetsang, avrillliu\}@hkust-gz.edu.cn}
}

\iclrfinalcopy 
\begin{document}

\maketitle

\begin{abstract}
Recent advances in AI-based speech synthesis have enabled highly realistic speech, increasing the importance of speech deepfake detection (SDD) in preventing misuse. While mainstream Self-Supervised Learning (SSL)-based detectors achieve strong performance, they suffer from poor generalization to unseen domains and often overlook fine-grained signal artifacts due to a bias towards global semantic consistency. In this paper, we conduct the first detailed empirical and visual analysis to validate these limitations explicitly. Our investigation reveals two critical architectural vulnerabilities: (1) a systemic failure to capture localized spoofing traces, and (2) a severe lack of adaptability to domain-driven shifts in SSL layer importance, rendering static aggregation strategies prone to overfitting. To address these vulnerabilities, we propose the Global-Local Adaptive Detector (GLAD). Specifically, to capture localized forgeries, GLAD employs a Hierarchical Global-Local (HGL) backbone that explicitly bridges the granularity gap by fusing global linguistic and acoustic features with fine-grained local signal details. To counter layer importance shifts in out-of-distribution (OOD) scenarios, we introduce a Hierarchical Adaptive Gating (HAG) mechanism that dynamically recalibrates layer-wise focus in a sample-specific manner. Finally, to address shortcut learning induced by environmental biases, we introduce SaniBoost, a composite data augmentation strategy for robust signal standardization and noise sanitization.
Extensive experiments demonstrate that GLAD significantly outperforms state-of-the-art methods, particularly on unseen domain cases.The code will be released upon publication.
\end{abstract}

\input{Sections/Introduction}

\input{Sections/RelatedWorks}

\input{Sections/Methods}
\input{Sections/Experiments}

\input{Sections/Conclusion}


\subsection*{AI use statement}

In this work, we used generative AI tools for assisting with translation. We have not used generative AI tools for helping develop theoretical models or conceptual frameworks, designing or providing feedback on research methodology or experiments, implementing methods, supporting qualitative and thematic data analysis, or interpreting results. Generating synthetic datasets, formulating mathematical claims, providing critical ingredients for proving mathematical claims, assisting in the writing of proofs, proposing or refining hypotheses, and cleaning or reformatting datasets are not applicable to this work. Additionally, we used generative AI tools for polishing the language and improving the readability of the manuscript. All AI-assisted content was reviewed by the authors, and the language-polished content was verified by two authors. We take responsibility for the final content of this work, including text, claims, or artifacts produced with the aid of generative AI.

\bibliography{iclr2027_conference}
\bibliographystyle{iclr2027_conference}
\newpage
\appendix

\begin{center}
    {\LARGE Appendix}
\end{center}

Due to space constraints in the main paper, this appendix provides additional experimental results, detailed implementation settings, visualization analyses, and extended ablation studies to complement the main findings. To ensure full reproducibility, we first provide detailed descriptions of the evaluation datasets and implementation settings of GLAD.
\section{Dataset Description and Implementation Details}
\label{app:data_impl}

\subsection{Evaluation Datasets}
\label{app:datasets}

\noindent \textit{ASVspoof 2019 LA (19LA)}~\cite{wang2020asvspoof}. This dataset contains TTS and VC attacks under logical access (LA) scenarios. We use the standard partition: 2,580 bona fide and 22,800 spoofed utterances for training, and 7,355 bona fide and 63,882 spoofed for evaluation.

\noindent \textit{ASVspoof 2021 LA (21LA) \& Deepfake (21DF)}~\cite{liu2023asvspoof}. The 21LA track inherits 19LA attacks but introduces varying transmission channels, comprising 166k spoofed samples. The 21DF track focuses on compressed deepfakes generated by over 100 unknown algorithms. It contains approximately 600k spoofed samples, presenting a rigorous test for unseen attacks.

\noindent \textit{Real-world Scenarios.} To evaluate robustness in the wild, we employ two datasets: (1) \textit{In-The-Wild (ITW)}~\cite{muller2022does}, containing 20.7h real and 17.2h deepfake speech collected from online platforms; and (2) \textit{Fake-or-Real (FoR)}~\cite{reimao2019dataset}, a balanced dataset designed to mitigate single-source generation bias.
\noindent \textit{Partial Spoof Scenarios}~\cite{zhang2022partialspoof}. To further evaluate performance under partial spoofing scenarios, we employ PartialSpoof, which is constructed from the ASVspoof 2019 LA corpus. Bona fide utterances are directly inherited from ASVspoof 2019 LA, while spoofed samples are generated by inserting spoofed segments into genuine speech recordings, resulting in partially manipulated utterances in which bona fide and spoofed regions coexist within the same audio sample. The official evaluation partition contains 7,355 bona fide utterances and 63,882 partially spoofed utterances.

\subsection{Training and Implementation Details}
\label{app:training}
\textbf{Input Processing.} All audio samples are resampled to 16 kHz and either cropped or zero-padded to a fixed length of 64,600 samples.
\textbf{Architecture Pipeline.} The overall framework consists of hierarchical dual-SSL feature extraction, cross-branch cross-attention interaction, adaptive layer selection/gating, raw-waveform CNN integration, attentive statistical pooling (ASP), and a final binary classification head.

\textbf{Signal-level CNN Branch.} The signal-level CNN branch extracts local spoofing representations directly from the raw waveform. Specifically, it first applies a SincConv1d front-end with 64 learnable band-pass filters for sub-band decomposition, followed by a 4-layer 1D convolutional stack for local temporal feature extraction. The first three convolutional layers use stride 2 for progressive downsampling, while the final layer adopts stride 1 to preserve temporal resolution. This branch is designed to capture fine-grained and band-related spoofing artifacts, thereby complementing the SSL encoders, which are generally biased toward high-level semantic and long-range contextual information. The subsequent linear projection and temporal interpolation are used only for feature alignment during later fusion stages, rather than being part of the feature extraction process itself.

\textbf{Augmentation Strategy.} For bona fide samples, 35\% are processed using the proposed clean-speech augmentation strategy (SaniBoost), while the remaining 65\% use RawBoost augmentation. All spoof samples are augmented using RawBoost.

\textbf{Hyperparameters.} Unless otherwise specified, the default random seed is set to 1234.

\textbf{Inference Time.} GLAD achieves an inference time of 21.08 ms/sample. Although slower than lightweight single-encoder models such as TCM (5.78 ms) and WaveSpec (14.31 ms), it still comfortably satisfies real-time processing requirements while delivering substantially stronger cross-domain detection performance.

\section{Additional Experimental Results}
\label{app:experiments}

\subsection{Performance Stability}
\begin{table*}[!t]
\centering
\caption{Performance stability comparison across different methods. Results are reported as mean $\pm$ standard deviation over multiple random seeds. EER values are reported in \%.}
\label{tab:seed_stability}

\small
\setlength{\tabcolsep}{4.5pt}
\renewcommand{\arraystretch}{1.08}

\resizebox{\textwidth}{!}{%
\begin{tabular}{lcccccc}
\toprule
\textbf{Method}
& \shortstack{\textbf{2021 LA}\\\textbf{EER$\downarrow$}}
& \shortstack{\textbf{2021 LA}\\\textbf{min-tDCF$\downarrow$}}
& \textbf{2021 DF$\downarrow$}
& \textbf{PartialSpoof$\downarrow$}
& \textbf{ITW$\downarrow$}
& \textbf{FoR$\downarrow$} \\
\midrule

AASIST
& $30.54 \pm 15.44$
& $0.8488 \pm 0.2615$
& $17.76 \pm 2.24$
& $35.92 \pm 4.53$
& $45.15 \pm 2.63$
& $32.77 \pm 12.37$ \\

W2V2-ST
& $3.93 \pm 3.02$
& $0.2896 \pm 0.0770$
& $4.61 \pm 3.05$
& $28.18 \pm 14.94$
& $11.78 \pm 6.00$
& $8.33 \pm 2.01$ \\

AMSDF
& $4.05 \pm 1.78$
& $0.2961 \pm 0.0483$
& $5.30 \pm 1.44$
& $14.74 \pm 0.81$
& $15.06 \pm 2.08$
& $18.40 \pm 7.26$ \\

XLSR-LGF
& $10.85 \pm 4.03$
& $0.4168 \pm 0.0679$
& $5.74 \pm 0.89$
& $9.37 \pm 0.22$
& $23.31 \pm 2.72$
& $28.19 \pm 5.99$ \\

XLSR-SLS
& $4.33 \pm 0.31$
& $0.2959 \pm 0.0087$
& $3.34 \pm 1.23$
& $12.11 \pm 4.04$
& $11.10 \pm 5.15$
& $11.83 \pm 6.29$ \\

XLSR-MultiConv
& $9.06 \pm 2.90$
& $0.4131 \pm 0.0732$
& $6.92 \pm 5.41$
& $9.83 \pm 3.00$
& $14.45 \pm 6.17$
& $25.60 \pm 2.75$ \\

STCA\&LMDC
& $3.27 \pm 1.41$
& $0.2745 \pm 0.0377$
& $9.43 \pm 2.87$
& $10.64 \pm 3.61$
& $20.26 \pm 1.85$
& $29.22 \pm 11.10$ \\

XLSR-Conformer\&TCM
& $4.18 \pm 0.53$
& $0.2994 \pm 0.0129$
& $2.68 \pm 0.08$
& $8.70 \pm 1.15$
& $10.41 \pm 1.12$
& $19.30 \pm 18.47$ \\

WaveSpec
& $15.62 \pm 1.49$
& $0.5037 \pm 0.0492$
& $18.90 \pm 2.34$
& $23.75 \pm 4.82$
& $26.41 \pm 2.59$
& $47.98 \pm 6.07$ \\

All-typed ADD
& $10.91 \pm 0.35$
& $0.4665 \pm 0.0192$
& $7.80 \pm 1.67$
& $11.00 \pm 1.87$
& $14.56 \pm 4.25$
& $16.21 \pm 9.32$ \\

WaveSP
& $14.21 \pm 6.49$
& $0.5601 \pm 0.1745$
& $15.63 \pm 10.12$
& $12.34 \pm 2.78$
& $27.78 \pm 18.24$
& $36.89 \pm 21.92$ \\

\midrule

\textbf{GLAD, 3 seeds}
& $\mathbf{2.70 \pm 0.74}$
& $\mathbf{0.2608 \pm 0.0186}$
& $\mathbf{1.53 \pm 0.29}$
& $\mathbf{7.17 \pm 0.86}$
& $\mathbf{5.91 \pm 0.61}$
& $\mathbf{1.59 \pm 0.70}$ \\

\bottomrule
\end{tabular}%
}

\end{table*}

\noindent
Although many prior ASVspoof  studies report only single seed performance, we think this is insufficient to establish robust superiority. To assess performance stability, we retrained GLAD across three different random seeds 
(909, 1234, 1235) to rule out potential ``lucky draw'' effects. 
The consistently low standard deviations across all datasets demonstrate that the proposed framework maintains stable and reliable performance under different random initializations, but several baselines exhibit considerably large variance. For instance, WaveSP has standard deviations of 10.12 points on 2021 DF dataset and 18.24 on ITW. By contrast, GLAD's corresponding standard deviations are 0.29 and 0.61, which shows in Table~\ref{tab:seed_stability}.
\subsection{Performance on 19LA}
\label{app:19la}
\begin{table}[!htbp]
\centering
\caption{Performance comparison on the ASVspoof 2019 LA evaluation set. Results are reported in terms of EER (\%) and min t-DCF. GLAD achieves SOTA performance, demonstrating superior robustness across most individual attack types (A07–A19). }
\label{tab:19la_comparison}
\resizebox{\textwidth}{!}{%
\begin{tabular}{l|ccccccccccccc|cc}
\toprule
\textbf{Method} & \textbf{A07$\downarrow$} & \textbf{A08$\downarrow$} & \textbf{A09$\downarrow$} & \textbf{A10$\downarrow$} & \textbf{A11$\downarrow$} & \textbf{A12$\downarrow$} & \textbf{A13$\downarrow$} & \textbf{A14$\downarrow$} & \textbf{A15$\downarrow$} & \textbf{A16$\downarrow$} & \textbf{A17$\downarrow$} & \textbf{A18$\downarrow$} & \textbf{A19$\downarrow$} & \textbf{EER$\downarrow$} & \textbf{min t-DCF$\downarrow$} \\ 
\midrule

TSSDN      & 1.43 & 0.75 & 0.02 & 1.75 & \textbf{0.06} & 0.18 & 0.06 & 0.11 & 2.05 & 1.25 & 6.01 & 1.14 & 1.44 & 1.62 & 0.0474 \\
RawGAT-ST  & 0.11 & 0.30 & 0.04 & \textbf{0.16} & 0.09 & 0.19 & 0.07 & 0.07 & 0.14 & 0.27 &  0.53  &\textbf{0.1} & 0.23 & 1.15 & 0.0373 \\

AASIST    & 0.80 & 0.44 & \textbf{0.00} & 1.06 & 0.16 & 0.31 & 0.91 & 0.1 & 0.15 & 0.65 & 0.72 & 1.52 & 3.40 & 0.62 & 1.13 \\
LC-Res18     & 0.05 & 0.26 & \textbf{0.00} & 0.26 & 0.13 & 0.17 & 0.18 & 0.10 & 0.18 & 0.10 & 2.48 & 0.40 & \textbf{0.17} & 0.80 & 0.0210 \\

W2V2-ST   & 0.06 & 0.06 & 0.02 & 0.40 & 0.10 & 0.14 & \textbf{0.00} & 0.06 & 0.24 & 0.06 & 0.37 & 0.84 & 0.35 & 0.25 & 0.0071 \\
AMSDF          & 0.03 & 0.05 & 0.03 & 0.52 & 0.38 & 0.23 & 0.02 & 0.05 & 0.33 & 0.12 & 0.30 & 0.47 & 0.41 & 0.31 & 0.0097 
\\
\textbf{GLAD (Ours)} & \textbf{0.00} & \textbf{0.04} & \textbf{0.00} & 0.18 & \textbf{0.06} & \textbf{0.06} & \textbf{0.00} & \textbf{0.00} & \textbf{0.00} & \textbf{0.02} & \textbf{0.28} & 0.57 & 0.31 & \textbf{0.14} & \textbf{0.0034} \\
\bottomrule
\end{tabular}%
}
\end{table}

As shown in Table~\ref{tab:19la_comparison}, GLAD establishes a new SOTA, achieving the lowest overall EER and min t-DCF. In particular, GLAD achieves the best performance on A07, A08, and A11–A16, demonstrating strong robustness across a wide range of spoofing attacks.

\subsection{Does larger parameters gurantee better performance?}
\begin{table*}[t]
\centering
\caption{
Comparison with large-scale and multi-backbone speech deepfake detectors.
Results are reported in EER (\%). ``--'' denotes results not reported in the corresponding work. Despite using substantially larger models or multiple pretrained backbones, existing methods do not consistently outperform GLAD across cross-domain evaluation benchmarks.
}
\label{tab:large_model_comparison}

\small
\setlength{\tabcolsep}{5pt}
\renewcommand{\arraystretch}{1.08}

\resizebox{\textwidth}{!}{%
\begin{tabular}{l l c c c c}
\toprule
\textbf{Model / Paper}
& \textbf{Parameters / Backbone}
& \textbf{21LA$\downarrow$}
& \textbf{21DF$\downarrow$}
& \textbf{ITW$\downarrow$}
& \textbf{FoR$\downarrow$} \\
\midrule

AWaveFormer~\cite{wang2025awaveformer}
& 634M, XLS-R + WavLM
& 2.33 & 3.63 & 10.25 & 5.15 \\

Towards Generalisable and Calibrated Audio Deepfake Detection~\cite{pascu2024towards}
& 2.16B, XLS-R-2B
& -- & -- & 7.20 & 6.30 \\

A Robust Audio Deepfake Detection System via Multi-view Feature~\cite{yang2024robust}
& 940M, XLS-R + WavLM + HuBERT
& 6.56 & -- & -- & -- \\

Comparative Analysis of ASR Methods for Speech Deepfake Detection~\cite{salvi2024comparative}
& 769M, Whisper Medium
& -- & 12.40 & 32.19 & 12.50 \\

Comparative Analysis of ASR Methods for Speech Deepfake Detection~\cite{salvi2024comparative}
& 1.55B, Whisper Large
& -- & 11.68 & 30.40 & 5.08 \\

ALLM4ADD~\cite{gu2025allm4add}
& 7.7B
& -- & -- & 26.99 & -- \\

\midrule
\textbf{GLAD (Ours)}
& \textbf{662.23M}
& \textbf{1.88}
& \textbf{1.31}
& \textbf{5.44}
& \textbf{1.10} \\

\bottomrule
\end{tabular}%
}
\end{table*}
In the section, we want to discuss whether large paramethers can induce a better performance? Table~\ref{tab:large_model_comparison} further examines whether the performance gains can be attributed simply to increased model capacity or the use of multiple pretrained backbones. The results suggest otherwise. AWaveFormer, which also combines XLS-R and WavLM and contains 634M parameters, yields EERs of 2.33\%, 3.63\%, 10.25\%, and 5.15\% on 21LA, 21DF, ITW, and FoR, respectively, while GLAD achieves substantially lower EERs of 1.88\%, 1.31\%, 5.44\%, and 1.10\% on the same benchmarks.Similarly, a 940M-parameter multi-view detector integrating XLS-R, WavLM, and HuBERT still reports an EER of 6.56\% on 21LA. Scaling the backbone alone also does not guarantee robust cross-domain generalization: Whisper Large (1.55B parameters) obtains 30.40\% EER on ITW, while  enlarge five times ALLM4ADD only reaches 26.99\%.
These comparisons indicate that simply increasing model capacity or stacking additional pretrained encoders is insufficient for robust speech deepfake detection. Instead, the gains of GLAD stem from explicitly modeling complementary global and local forensic cues and adaptively recalibrating their contributions across layers and samples.

\subsection{Further Performance Analysis on A18 and A19}
GLAD shows slightly higher EER on A18 and A19 compared to several handcrafted-feature-based baselines. This is mainly because A18 and A19 are traditional statistical voice conversion attacks, whose artifacts primarily manifest in low-level cepstral distortions, over-smoothed spectral envelopes, and unnatural excitation–filter coupling patterns. In contrast, SSL representations inherently prioritize high-level semantic and speaker-related information, making them naturally less sensitive to such localized low-level vocoder artifacts.

To further examine this phenomenon, we repeated the experiments using different random seeds. We observe that the performance on A18 and A19 exhibits certain fluctuations across runs, indicating that these low-level statistical artifacts remain challenging for SSL-based architectures. Nevertheless, GLAD consistently maintains strong and stable performance, achieving averaged EERs of 0.46 ± 0.035 on A18 and 0.30 ± 0.039 on A19.
More importantly, GLAD consistently outperforms representative SSL-based baselines. For example, W2V2-ST and AMSDF achieve EERs of (0.84, 0.35) and (0.47, 0.41) on A18 and A19, respectively, while GLAD further reduces them to (0.43, 0.28). These results demonstrate that the proposed framework effectively alleviates the intrinsic weakness of SSL representations on legacy vocoder-based spoofing attacks.
However, compared with performance on conventional spoofing attacks, GLAD still shows noticeable limitations on A18 and A19. This suggests that shallow statistical artifacts and excitation-related distortions remain difficult for current SSL-based representations to fully capture. In future work, we plan to further investigate how to improve SSL models for identifying both conventional and low-level spoofing attacks more effectively.

Overall, GLAD achieves consistently strong within-domain performance across nearly all attack types and demonstrates superior robustness against modern neural-based spoofing attacks, which represent the dominant trend in contemporary deepfake generation. However, the remaining performance gap on traditional low-level attacks such as A18 and A19 also suggests that current SSL-based representations still have limitations in modeling subtle statistical spoofing artifacts.

\section{Quantitative Attribution Alignment on Short Partial Spoofs}
\label{app:local_attribution}
To complement the motivating saliency visualization, we further conduct a quantitative analysis to examine whether the attribution maps produced by GLAD are spatially aligned with annotated short spoofed regions. Specifically, we select all PartialSpoof utterances whose total annotated spoof duration does not exceed 0.20\,s and group them according to the number of annotated spoof segments. GLAD is then compared with SLS using the same attribution extraction and evaluation protocol.

We report three metrics: Frame AUPRC, IoU@GT, and Segment Recall. The results are reported in Table~\ref{tab:local_attribution}. Compared with SLS, GLAD improves Frame AUPRC from 0.2181 to 0.2811, IoU@GT from 0.1124 to 0.1565, and Segment Recall from 0.2674 to 0.4000, corresponding to relative improvements of 28.89\%, 39.23\%, and 49.59\%, respectively, which means that GLAD not only improves utterance-level detection performance, but also produces attribution maps that are more closely aligned with short, localized spoofing regions.

\begin{table}[htbp]
\centering
\caption{Attribution alignment on PartialSpoof utterances.}
\label{tab:local_attribution}
\small
\setlength{\tabcolsep}{4pt}
\begin{tabular}{crrrrrr}
\toprule
& \multicolumn{2}{c}{Frame AUPRC} & \multicolumn{2}{c}{IoU@GT} & \multicolumn{2}{c}{Segment Recall} \\
\cmidrule(lr){2-3}\cmidrule(lr){4-5}\cmidrule(lr){6-7}
\# segments & XLS-R-SLS & GLAD & XLS-R-SLS & GLAD & XLS-R-SLS & GLAD \\
\midrule
1  & 0.1734 & 0.2034 & 0.0822 & 0.0898 & 0.2360 & 0.3202 \\
2  & 0.3025 & 0.3894 & 0.1636 & 0.2346 & 0.4022 & 0.5858 \\
3  & 0.2224 & 0.2786 & 0.1135 & 0.1458 & 0.2545 & 0.3737 \\
4  & 0.2072 & 0.2553 & 0.1050 & 0.1340 & 0.2232 & 0.3406 \\
5  & 0.1590 & 0.2109 & 0.0801 & 0.1086 & 0.1743 & 0.2770 \\
6  & 0.1359 & 0.1949 & 0.0612 & 0.1071 & 0.1537 & 0.2799 \\
7  & 0.1270 & 0.1698 & 0.0597 & 0.0870 & 0.1414 & 0.2158 \\
8  & 0.1060 & 0.1714 & 0.0494 & 0.0942 & 0.1324 & 0.2194 \\
9  & 0.0980 & 0.1451 & 0.0428 & 0.0708 & 0.0862 & 0.1678 \\
10 & 0.1010 & 0.1494 & 0.0405 & 0.0760 & 0.0929 & 0.1964 \\
\midrule
Overall & 0.2181 & \textbf{0.2811} & 0.1124 & \textbf{0.1565} & 0.2674 & \textbf{0.4000} \\
\bottomrule
\end{tabular}
\end{table}

As shown in Table~\ref{tab:local_attribution}, GLAD improves Frame AUPRC
from 0.2181 to 0.2811, IoU@GT from 0.1124 to 0.1565, and Segment Recall
from 0.2674 to 0.4000. The reported relative improvements are 28.86\%,
39.26\%, and 49.59\%, respectively. Improvements are observed for all three
metrics in every segment-count group from one to ten, including utterances
containing multiple short, fragmented manipulations. These results extend
the single illustrative example and support the observation that the complete
GLAD detector produces attributions better aligned with localized spoof
evidence than the evaluated XLS-R-SLS baseline.

\section{Additional Visualization Studies}
\label{app:visualization}
\begin{figure*}[!t]
    \centering
    \includegraphics[width=\linewidth]{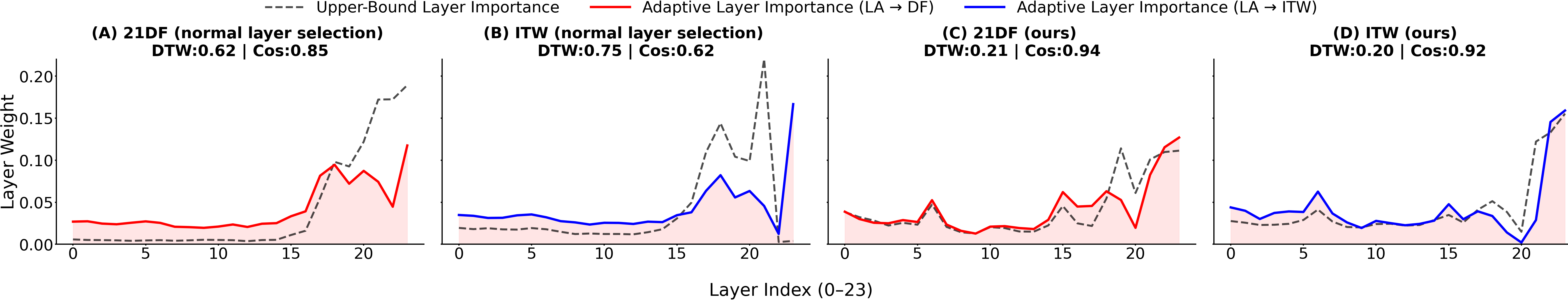}
    \caption{
    Visualization of layer importance distributions under different layer selection strategies across multiple random seeds. Panels (A–B) show the results obtained using Sensitive Layer Selection (SLS)~\cite{zhang2024audio}, while Panels (C–D) correspond to the proposed Hierarchical Adaptive Gating (HAG) mechanism. The solid curves (Red for 21DF and Blue for ITW) represent the adaptive layer importance learned by models trained on the source domain (19LA), whereas the grey dashed curves denote the oracle layer importance distributions derived from in-domain training (upper bound). The shaded regions indicate the standard deviation across different random seeds, reflecting the stability of the learned layer preferences. Compared with SLS, the proposed HAG in Panels (C–D) achieves substantially lower DTW distances and higher cosine similarities with the oracle distributions, demonstrating more stable and accurate cross-domain layer adaptation behavior.
    }
    \label{fig:xlsr_compare}
\end{figure*}
To further verify that the proposed layer selection and gating mechanisms are stable rather than being caused by specific backbone effects, we conduct additional visualization studies under different backbone settings. Specifically, we replace the SSL backbone with XLS-R and retrain the model using seed 1234 while keeping all other configurations unchanged.

Figure~\ref{fig:xlsr_compare} illustrates the resulting adaptive layer importance distributions under different backbone configurations. We observe that the overall layer adaptation trends remain highly consistent across different backbone configurations. In particular, the adaptive layer importance learned by GLAD consistently aligns well with the corresponding upper-bound layer importance distributions, especially on the proposed model variants shown in Figure~(C) and Figure~(D). Compared with the baseline settings in Figure~(A) and Figure~(B), the proposed adaptive gating mechanism achieves significantly lower DTW distances and higher cosine similarities, indicating more stable and accurate layer selection behavior.

These observations suggest that the proposed hierarchical adaptive gating mechanism does not rely on lucky initialization or seed-specific optimization trajectories. Instead, it consistently learns meaningful layer importance patterns that generalize across datasets and experimental settings.

\section{Additional Analysis of Each Module}
\subsection{Functional Analysis of HAG Gating}
\label{app:hag_intervention}
To investigate whether HAG benefits from utterance-conditioned recalibration rather than a fixed domain-level preference, we perform
a controlled gate intervention using the same trained GLAD checkpoint. We compare two inference conditions. \emph{Dynamic} uses each
utterance's own layer, stream, and local gates, whereas \emph{Target Mean} replaces these gates with their corresponding mean values estimated from the target domain.
To avoid self inclusion, we split the data into two folds and perform cross fitting, while keeping all other model parameters unchanged. We define the performance degradation caused by replacing the dynamic gates with averages computed over the target domain as
\begin{equation}
\Delta \mathrm{EER}
=
\mathrm{EER}_{\mathrm{TargetMean}}
-
\mathrm{EER}_{\mathrm{Dynamic}}.
\end{equation}
We estimate 95\% confidence intervals using 1,000 class-stratified
paired bootstrap resamples with seed 1234. In this section, 21LA includes the eval, progress, and hidden subsets and
is therefore not directly comparable with the eval-only result in the
main benchmark.

\begin{table}[t]
\centering
\caption{
Gate intervention using the same trained GLAD checkpoint.
}
\label{tab:hag_intervention}

\small
\setlength{\tabcolsep}{5pt}

\begin{tabular}{lccc}
\toprule
\textbf{Dataset}
& \textbf{Dynamic}
& \textbf{Target-domain Mean}
& \textbf{$\Delta$EER [95\% CI]} \\
\midrule

21LA
& \textbf{5.3030}
& 5.7285
& +0.4255 [0.3268, 0.5223] \\

21DF
& \textbf{1.3108}
& 1.4709
& +0.1601 [0.1277, 0.2048] \\

ITW
& \textbf{5.4418}
& 5.6195
& +0.1777 [0.0592, 0.2877] \\

\bottomrule
\end{tabular}
\end{table}

As shown in Table~\ref{tab:hag_intervention}, Dynamic consistently outperforms target domain mean on all three datasets, with all
bootstrap confidence intervals lying above zero. Importantly, the static condition already uses statistics estimated from the target domain, yet averaging the gates still degrades performance. This suggests that HAG benefits from preserving utterance-specific gate adaptation rather than relying on a single domain-level weighting pattern.

To further examine whether such gate variation merely reflects the difference between bona fide and spoof classes, we report the
proportion of gate variation that remains within the two classes.

\begin{table}[t]
\centering
\caption{
Proportion of dynamic-gate variation remaining within bona fide and spoof classes. Higher values indicate that a larger fraction of gate variation occurs among utterances within the same class.
}
\label{tab:hag_variation}

\small
\setlength{\tabcolsep}{7pt}

\begin{tabular}{lccc}
\toprule
\textbf{Dataset}
& \textbf{Layer (\%)}
& \textbf{Stream (\%)}
& \textbf{Local (\%)} \\
\midrule

21LA & 61.2 & 55.9 & 98.6 \\
21DF & 66.0 & 74.0 & 99.8 \\
ITW  & 82.2 & 65.0 & 98.1 \\

\bottomrule
\end{tabular}
\end{table}

As shown in Table~\ref{tab:hag_variation}, 55.9--82.2\% of the variation in the layer gates and stream gates occurs within classes, while more than 98\% of the variation in the local gates occurs within classes across all three datasets. These findings further show that HAG dynamically adjusts feature weights for each input, rather than relying on fixed class or domain level weights.

\subsection{Signal and Representation Analysis of SaniBoost}
In this section, we investigate what acoustic characteristics are modified by SaniBoost and what effects these modifications may introduce. We first construct a controlled subset of bona fide utterances from the 19LA training set. The training set contains 20 speakers; for each speaker, we randomly select 25 bona fide utterances using a fixed random seed of 1234, resulting in a total of 500 utterances. Audio duration, VCTK speaker identifiers, and reference transcriptions are retrieved only after the sampling procedure and are not used as selection criteria.

\textbf{Spectral and amplitude modifications.} To quantify the signal-level changes introduced by SaniBoost, we measure the energy change in four frequency bands, the shift in spectral centroid, and the change in RMS amplitude. Band-wise energy and RMS changes are reported relative to the original waveform. We additionally compute, for each utterance, the proportion of samples set to zero by the noise-gating operation. For all measurements, we report the median together with the 25th and 75th percentiles across the 500 utterances.Table~\ref{tab:saniboost_signal} shows the results.

\begin{table}[t]
\centering
\caption{
Paired signal measurements on the fixed, speaker-balanced subset.
}
\label{tab:saniboost_signal}

\small
\setlength{\tabcolsep}{5pt}
\renewcommand{\arraystretch}{1.08}

\begin{tabular}{lc}
\toprule
\textbf{Measurement}
& \textbf{Median [25th, 75th percentile]} \\
\midrule

Energy change below 500\,Hz
& $-40.07\;[-41.92,-38.64]$\,dB \\

Energy change, 500--2000\,Hz
& $-13.04\;[-15.13,-10.97]$\,dB \\

Energy change, 2000--4000\,Hz
& $+2.56\;[+2.31,+2.77]$\,dB \\

Energy change, 4000--8000\,Hz
& $+3.43\;[+3.36,+3.47]$\,dB \\

Spectral centroid change
& $+2.63\;[+2.15,+3.19]$\,kHz \\

RMS amplitude change
& $-10.71\;[-13.16,-8.64]$\,dB \\

Waveform samples set to zero by gating
& $77.23\;[72.94,81.39]$\% \\

\bottomrule
\end{tabular}
\end{table}

Median energy decreases by 40.07\,dB below 500\,Hz and by 13.04\,dB in the 500--2000\,Hz band, while the changes in the two higher-frequency bands are positive.
The spectral centroid increases by a median of 2.63\,kHz, the RMS amplitude decreases by 10.71\,dB, and the median fraction of samples set to zero is 77.23\%. These results shows that SaniBoost produces a substantial acoustic transformation rather than a small background-noise reduction.

\textbf{Objective intelligibility and ASR transcription accuracy.}
We evaluate Short-Time Objective Intelligibility (STOI)~\cite{taal2010short} using the original waveform as the reference. Across the selected utterances, the median STOI is 0.808, with an interquartile range of $[0.769, 0.835]$.

We further assess transcription consistency using Whisper large-v3~\cite{radford2023robust}, which transcribes both the original and processed waveforms under identical settings. The reference and recognized transcripts are normalized using the Whisper text-normalization procedure before evaluation with jiwer. We report both Word Error Rate (WER) and Character Error Rate (CER). In addition, a 95\% confidence interval for the change in corpus-level WER is estimated using 20,000 paired bootstrap resamples. The results are shown in Table~\ref{tab:saniboost_asr}.

\begin{table}[t]
\centering
\caption{
Whisper large-v3 transcription analysis. WER and CER are corpus-level
percentages. The WER difference is computed before rounding, and its
brackets indicate a paired-bootstrap 95\% confidence interval.
The final three rows compare utterance-level WER before and after
sanitization.
}
\label{tab:saniboost_asr}

\small
\setlength{\tabcolsep}{5pt}
\renewcommand{\arraystretch}{1.08}

\resizebox{\linewidth}{!}{%
\begin{tabular}{lccc}
\toprule
\textbf{Measurement}
& \textbf{Original}
& \textbf{Forced SaniBoost}
& \textbf{Change} \\
\midrule

Word errors / reference words
& 48/3523
& 132/3523
& $+84$ errors \\

Corpus WER (\%)
& 1.362
& 3.747
& $+2.384$  \\

Corpus CER (\%)
& 0.611
& 1.957
& $+1.346$  \\

\midrule

95\% CI for the WER increase
& \multicolumn{3}{c}{$[+1.678,+3.126]$ } \\

Unchanged utterance-level WER
& \multicolumn{3}{c}{431/500 (86.2\%)} \\

Increased utterance-level WER
& \multicolumn{3}{c}{64/500 (12.8\%)} \\

Decreased utterance-level WER
& \multicolumn{3}{c}{5/500 (1.0\%)} \\

\bottomrule
\end{tabular}%
}
\end{table}

The results show that the processed speech remains largely transcribable, with a corpus-level WER of 3.747\%. Compared with the original WER of 1.362\%, SaniBoost increases WER by 2.384 \%, with the 95\% paired bootstrap confidence interval remaining above zero. This corresponds to 84 additional word errors over 3,523 reference words. At the utterance level, 431 samples show unchanged WER, while 64 deteriorate and five improve, which means that  SaniBoost introduces substantial acoustic modifications while largely preserving linguistic content, although with a measurable transcription cost.

\textbf{Internal representation consistency.} We further examine how SaniBoost affects the internal representations learned by GLAD. Here, $p$ denotes the application probability of Adversarial Bona Fide Purification (ABFP) in SaniBoost.

For each checkpoint, we feed the 500 paired original and sanitized utterances through GLAD and extract representations at different stages. Intermediate representations are averaged over the remaining sequence dimensions to obtain one vector per utterance, while the final representation is taken after attentive statistics pooling and projection. Representation similarity is evaluated using two metrics: cosine similarity and linear Centered Kernel Alignment (CKA)~\cite{kornblith2019similarity}. Table~\ref{tab:saniboost_representation} reports the results for checkpoints trained with $p=0$ and $p=0.35$.

\begin{table}[t]
\centering
\caption{
Representation agreement under the same forced-sanitization protocol.
The two training settings differ in their ABFP probability, $p=0$ or $p=0.35$.
}
\label{tab:saniboost_representation}

\small
\setlength{\tabcolsep}{5pt}
\renewcommand{\arraystretch}{1.10}

\begin{tabular}{lcccc}
\toprule
& \multicolumn{2}{c}{\textbf{Mean Paired Cosine}}
& \multicolumn{2}{c}{\textbf{Linear CKA}} \\
\cmidrule(lr){2-3}
\cmidrule(lr){4-5}

\textbf{Stage}
& $p=0$
& $p=0.35$
& $p=0$
& $p=0.35$ \\
\midrule

WavLM
& 0.8965 & 0.8858
& 0.7522 & 0.6480 \\

XLS-R
& 0.6896 & 0.8704
& 0.0877 & 0.6764 \\

Cross-SSL
& 0.7325 & 0.8724
& 0.1327 & 0.6597 \\

CNN
& 0.9783 & 0.9913
& 0.1794 & 0.1787 \\

Multi-stream fusion
& 0.3095 & 0.9096
& 0.0407 & 0.3705 \\

Pooling and projection
& 0.5211 & 0.9854
& 0.0270 & 0.3702 \\

\bottomrule
\end{tabular}
\end{table}

The largest gains in representation agreement occur at the later fusion stages. With \(p=0.35\), the mean paired cosine increases from 0.3095 to 0.9096 after multi-stream fusion and from 0.5211 to 0.9854 after pooling and projection. Linear CKA follows the same trend, suggesting that the improvement reflects not only stronger pairwise alignment but also greater preservation of the global representation geometry across utterances. Notably, the effect is not uniform across individual encoders: WavLM shows a slight decrease in agreement, whereas XLS-R and Cross-SSL become substantially more aligned. This pattern suggests that SaniBoost encourages the middle and later stages of the model to become less sensitive to sanitization-induced variations.

\textbf{Training exposure and classification behavior of ABFP.} Furthermore, we compare the predictions of two trained checkpoints (\(p=0\) and \(p=0.35\)) on the same data under two input conditions. The original condition refers to inference on the unmodified waveform without any data augmentation, whereas the sanitized condition applies ABFP to every waveform before inference. A bona fide probability threshold of 0.5 is used for binary classification. The results are reported in Table~\ref{tab:saniboost_training_control}.

\begin{table}[t]
\centering
\caption{
Prediction behavior under forced sanitization. All 500 inputs are
bona fide. Both checkpoints use a bona fide probability threshold of 0.5.
}
\label{tab:saniboost_training_control}

\small
\setlength{\tabcolsep}{5pt}
\renewcommand{\arraystretch}{1.08}

\begin{tabular}{lcc}
\toprule
\textbf{Measurement}
& $\boldsymbol{p=0}$
& $\boldsymbol{p=0.35}$ \\
\midrule

Original inputs classified as bona fide
& 498/500
& 499/500 \\

Sanitized inputs classified as bona fide
& 278/500
& 500/500 \\

Conditional bona fide retention
& 277/498
& 499/499 \\

All predictions unchanged
& 278/500
& 499/500 \\

\midrule

Spoof $\rightarrow$ spoof
& 1 & 0 \\

Spoof $\rightarrow$ bona fide
& 1 & 1 \\

Bona fide $\rightarrow$ spoof
& 221 & 0 \\

Bona fide $\rightarrow$ bona fide
& 277 & 499 \\

\bottomrule
\end{tabular}
\end{table}
This decision-level analysis is consistent with the representation-level results. The \(p=0.35\) checkpoint preserves nearly perfect classification on the original inputs (499/500) while remaining fully stable under the ABFP condition, with all 499 originally bona fide predictions retained after sanitization. In contrast, the \(p=0\) checkpoint performs similarly on the original inputs but exhibits substantial prediction instability after sanitization, with 221 bona fide predictions switching to spoof. These results indicate that incorporating ABFP during training preserves the model's discriminative capability on unmodified inputs while making its decisions substantially more stable under the corresponding sanitization transformation.

\textbf{Independent speaker representation consistency.}
We additionally assess whether speaker representations remain consistent after sanitization using a separate pretrained speaker verification model, WavLM-Base-Plus for Speaker Verification~\cite{chen2022wavlm}. For each of the 500 original--sanitized pairs, we extract utterance-level speaker embeddings and apply L2 normalization.

For normalized embeddings $\mathbf{e}_i^{\mathrm{orig}}$ and $\mathbf{e}_i^{\mathrm{san}}$, the paired cosine similarity is defined as
\begin{equation}
c_i^{\mathrm{spk}} =
\left(\mathbf{e}_i^{\mathrm{orig}}\right)^{\top}
\mathbf{e}_i^{\mathrm{san}}.
\end{equation}

We summarize the results over the 500 pairs using the median and interquartile interval, with the mean reported as a complementary statistic.

\begin{table}[t]
\centering
\caption{
Speaker representation similarity between the original and forcibly
sanitized utterances.
}
\label{tab:saniboost_speaker}

\small
\setlength{\tabcolsep}{7pt}
\renewcommand{\arraystretch}{1.08}

\begin{tabular}{lc}
\toprule
\textbf{Measurement} & \textbf{Result} \\
\midrule

Median paired speaker cosine
& 0.884 \\

Interquartile interval
& $[0.827,0.919]$ \\

Mean paired speaker cosine
& 0.860 \\

\bottomrule
\end{tabular}
\end{table}

As shown in Table~\ref{tab:saniboost_speaker}, the median paired cosine similarity is 0.884, with an interquartile interval of $[0.827, 0.919]$ and a mean of 0.860. Despite the substantial spectral and amplitude modifications introduced by SaniBoost, the original and sanitized utterances retain considerable similarity in the representation space of an independently pretrained speaker verification model.

In conclusion, At the signal level, SaniBoost introduces substantial spectral and amplitude changes while largely preserving speech intelligibility and transcribability, although sanitization incurs a measurable increase in ASR errors. The independent speaker analysis further shows that considerable speaker-related information is retained after transformation. At the representation level, models trained with SaniBoost exhibit greater consistency in fused and final representations between original and sanitized speech, together with fewer bona fide errors.

\section{Paired Robustness Evaluation for SaniBoost}

We further evaluate the contribution of the asymmetric clean branch in SaniBoost to detection robustness under a specified class-conditional acoustic transformation protocal. To this end, we train three models: (1) No augmentation,
(2) Without ABFP, and (3) Full SaniBoost. All models follow the same
hyperparameter selection protocol as GLAD. For each of 21LA, 21DF, and
ITW, we randomly sample 5,000 bona fide and 5,000 spoof utterances using
a fixed seed of 1234.

Let $C(x)$ denote the bona fide sanitization intervention, implemented
using the same 2,000~Hz high-pass operation adopted during training, and
let $N_\rho(x)$ denote spoof speech corrupted with environmental noise
at an SNR of $\rho\in\{10,5,0\}$~dB. We construct a \emph{conflict set}
by applying $C(x)$ to bona fide utterances and $N_\rho(x)$ to spoof
utterances while preserving their original class labels. The original
and conflict sets contain the same utterances, allowing a paired
comparison before and after the joint intervention.

For model $m$, we define the decision score as
\begin{equation}
s_m(x)=\log p_m(\mathrm{spoof}\mid x)
-\log p_m(\mathrm{bona\ fide}\mid x),
\label{eq:sb_score}
\end{equation}
where higher scores indicate stronger evidence of spoofing. We evaluate
performance using EER, computed by independently sweeping the decision
threshold for each model on the original subset and on the conflict
set at each SNR.

\paragraph{Joint Transformation Robustness.}
We compare detection performance before and after jointly applying sanitization to bona fide utterances and additive noise to spoof utterances. For each model and SNR, we quantify the EER change relative to the original subset as
\begin{equation}
\Delta E_m(\rho)
=
\mathrm{EER}_{m,\mathrm{conflict},\rho}
-
\mathrm{EER}_{m,\mathrm{original}},
\label{eq:sb_delta_eer}
\end{equation}
where a positive value indicates degradation and a negative value
indicates improvement. EER is reported as a percentage, and
$\Delta$EER is reported in percentage points.
Table~\ref{tab:sb_conflict_5db} compares the original and conflict
results at 5~dB, while Table~\ref{tab:sb_conflict_all_snr} reports
results across all tested SNRs.

No augmentation suffers severe degradation under the joint
intervention, while Without ABFP shows smaller but consistently
positive EER changes. In contrast, Full SaniBoost maintains low EER
across all three datasets and all tested SNRs, with negative
$\Delta$EER in all nine settings. At 5~dB, its EER decreases from
1.81\% to 0.02\% on 21LA, from 1.20\% to 0.04\% on 21DF, and from
7.69\% to 2.45\% on ITW. These results show that fully Saniboost achieves lower EER than both comparison models under the specified class-conditional transformations across all datasets and tested SNRS.

\begin{table}[htbp]
\centering
\caption{EER before and after the joint intervention at 5~dB SNR.
The conflict set contains sanitized bona fide utterances and spoof
utterances corrupted with environmental noise. EER is reported in
\%, and $\Delta$EER in percentage points.}
\label{tab:sb_conflict_5db}
\small
\setlength{\tabcolsep}{5pt}
\renewcommand{\arraystretch}{1.10}
\begin{tabular}{llrrr}
\toprule
\textbf{Dataset} & \textbf{Training condition}
& \shortstack{\textbf{Original}\\\textbf{EER}}
& \shortstack{\textbf{Conflict}\\\textbf{EER}}
& \textbf{$\Delta$EER} \\
\midrule
21LA & No augmentation & 4.50 & 71.44 & $+66.94$ \\
 & Without ABFP & 2.41 & 11.52 & $+9.11$ \\
 & Full SaniBoost & 1.81 & 0.02 & $-1.79$ \\
\midrule
21DF & No augmentation & 2.41 & 64.59 & $+62.18$ \\
 & Without ABFP & 1.96 & 23.86 & $+21.90$ \\
 & Full SaniBoost & 1.20 & 0.04 & $-1.16$ \\
\midrule
ITW & No augmentation & 8.30 & 61.30 & $+53.00$ \\
 & Without ABFP & 8.91 & 23.21 & $+14.30$ \\
 & Full SaniBoost & 7.69 & 2.45 & $-5.24$ \\
\bottomrule
\end{tabular}
\end{table}

\begin{table}[!t]
\centering
\caption{EER on the conflict set across different noise levels.
EER is reported in \%, and $\Delta$EER denotes the change from the
corresponding original subset in percentage points.}
\label{tab:sb_conflict_all_snr}
\small
\setlength{\tabcolsep}{8pt}
\renewcommand{\arraystretch}{1.10}
\begin{tabular}{lrrr}
\toprule
\textbf{Training condition}
& \textbf{SNR (dB)}
& \textbf{EER}
& \textbf{$\Delta$EER} \\
\midrule
\multicolumn{4}{l}{\textbf{21LA}} \\
No augmentation & 10 & 63.43 & $+58.93$ \\
 & 5 & 71.44 & $+66.94$ \\
 & 0 & 74.46 & $+69.96$ \\
Without ABFP & 10 & 10.15 & $+7.74$ \\
 & 5 & 11.52 & $+9.11$ \\
 & 0 & 11.30 & $+8.89$ \\
Full SaniBoost & 10 & 0.10 & $-1.71$ \\
 & 5 & 0.02 & $-1.79$ \\
 & 0 & 0.02 & $-1.79$ \\
\midrule
\multicolumn{4}{l}{\textbf{21DF}} \\
No augmentation & 10 & 51.45 & $+49.04$ \\
 & 5 & 64.59 & $+62.18$ \\
 & 0 & 73.47 & $+71.06$ \\
Without ABFP & 10 & 19.63 & $+17.67$ \\
 & 5 & 23.86 & $+21.90$ \\
 & 0 & 23.84 & $+21.88$ \\
Full SaniBoost & 10 & 0.07 & $-1.13$ \\
 & 5 & 0.04 & $-1.16$ \\
 & 0 & 0.01 & $-1.19$ \\
\midrule
\multicolumn{4}{l}{\textbf{ITW}} \\
No augmentation & 10 & 51.44 & $+43.14$ \\
 & 5 & 61.30 & $+53.00$ \\
 & 0 & 67.46 & $+59.16$ \\
Without ABFP & 10 & 20.36 & $+11.45$ \\
 & 5 & 23.21 & $+14.30$ \\
 & 0 & 23.45 & $+14.54$ \\
Full SaniBoost & 10 & 2.23 & $-5.46$ \\
 & 5 & 2.45 & $-5.24$ \\
 & 0 & 2.36 & $-5.33$ \\
\bottomrule
\end{tabular}
\end{table}

\paragraph{Paired Statistical Analysis.}
We further evaluate whether the reductions in EER change introduced
by ABFP are consistent across evaluation samples using 10,000 paired
bootstrap resamples. For each replicate, utterances are resampled
with replacement separately within each class. The same sampled
indices are reused across Full SaniBoost and Without ABFP and across
the original and conflict conditions. Each replicate recomputes
both models' EER changes and their paired contrast,
\begin{equation}
\theta(\rho)
=
\Delta E_{\mathrm{Full}}(\rho)
-
\Delta E_{\mathrm{Without\ ABFP}}(\rho).
\label{eq:sb_paired_contrast}
\end{equation}
A negative contrast indicates a smaller EER change for Full
SaniBoost relative to Without ABFP.

As shown in Table~\ref{tab:sb_paired_bootstrap}, the observed
contrasts at 5~dB are $-10.90$, $-23.06$, and $-19.54$ percentage
points on 21LA, 21DF, and ITW, respectively. The corresponding
95\% confidence intervals are entirely below zero. This shows
consistent reductions in EER change relative to the variant
without ABFP across bootstrap resamples.

\begin{table}[!t]
\centering
\caption{Full SaniBoost minus Without ABFP contrasts in
$\Delta$EER. All values are in percentage points. Confidence
intervals are obtained from 10,000 paired bootstrap resamples.
This setting uses 5~dB SNR.}
\label{tab:sb_paired_bootstrap}
\small
\setlength{\tabcolsep}{8pt}
\renewcommand{\arraystretch}{1.10}
\begin{tabular}{lrc}
\toprule
\textbf{Dataset}
& \shortstack{\textbf{Observed}\\\textbf{$\Delta$EER contrast}}
& \textbf{95\% CI} \\
\midrule
21LA & $-10.90$ & $[-11.56, -10.24]$ \\
21DF & $-23.06$ & $[-23.86, -22.14]$ \\
ITW & $-19.54$ & $[-20.58, -18.66]$ \\
\bottomrule
\end{tabular}
\end{table}

\section{Additional Ablation Studies}
\label{app:ablation}
To further validate the necessity of each component design, we conduct additional ablation studies on individual modules and further investigate different attention interaction strategies.

\begin{table}[t!]                                        
\centering
\caption{Ablation study on the internal core components of GLAD on 21LA, 21DF, and ITW datasets (EER \%). The Adaptive S-w Gating denotes the Stream-wise Gating, and the Refined L-w Aggregation denotes the Refined Layer-wise Aggregation. The L-A Cross-Attention denotes the Linguistic-Acoustic Cross-Attention. The ABFP means adversarial bona fide purification. Asymmetric ABFP means using ABFP adversarial purification for both bona fide and spoof.}

\label{tab:ablation_submodules}

\footnotesize
\setlength{\tabcolsep}{5pt}
\renewcommand{\arraystretch}{0.92}

\begin{tabular}{p{0.73\linewidth}ccc}
\toprule
\textbf{Method / Variant} 
& \textbf{21LA$\downarrow$} 
& \textbf{21DF$\downarrow$} 
& \textbf{ITW$\downarrow$} \\
\midrule

w/o L-A Cross-Attention & 3.82 & 12.31 & 12.82 \\
w/o Multi-Granularity Fusion & 2.21 & 2.05 & 8.33 \\

\midrule
w/o Adaptive S-w Gating & 2.26 & 1.71 & 7.67 \\
w/o Refined L-w Aggregation & 4.70 & 1.96 & 6.14 \\
w/o Global-Local Modulation & 2.37 & 1.66 & 6.19 \\

\midrule
w/ RawBoost & 3.64 & 1.90 & 7.80 \\
w/o asymmetric ABFP & 2.80 & 1.44 & 6.10 \\
w/o ABFP & 2.47 & 2.10 & 7.05 \\

\midrule
w/ One-way XLS-R\&WavLM attention & 1.57 & 3.36 & 6.75 \\
w/ Bidirectional CNN\&XLS-R attention & 2.00 & 1.63 & 6.10 \\

\midrule
\textbf{GLAD} & 1.88 & 1.31 & 5.44 \\
\bottomrule
\end{tabular}
\vspace{-1.0em}
\end{table}
\subsection{Additional Component-wise Ablation Studies}
\textbf{Impact of HGL Components.} As shown in the first part of Table~\ref{tab:ablation_submodules}, replacing \textit{ Linguistic-Acoustic Cross-Attention} with additive fusion significantly degrades performance, indicating that simple addition fails to model complex cross-modal dynamics. Likewise, substituting the attentive \textit{Multi-Granularity Fusion} with rigid concatenation leads to performance drops, confirming that attention-based integration provides a more informative representation than simple feature stacking.

\noindent \textbf{Impact of HAG components,} As shown in Table~\ref{tab:ablation_submodules}. Removing \textit{Adaptive Stream-wise gating} or \textit{Refined Layer-wise Aggregation} or \textit{Global-Local Modulation} degrades performance, validating its utility in filtering representations.

\noindent \textbf{Impact of Saniboost Components.}
Replacing SaniBoost with the RawBoost method impairs robustness, notably on 21LA (EER: 1.88\% $\to$ 3.64\%). Moreover, applying adversarial purification to both bona fide and spoof samples, or disabling \textit{adversarial bona fide purification} entirely, degrades performance which support the effectiveness of selective bona fide purification within SaniBoost.

\textbf{Impact of Attention Weight.} Using a one-way attention design (i.e., using XLS-R as query and WavLM as key/value) causes substantial performance drops on 21DF and ITW, despite a slight improvement on 21LA. This suggests that bidirectional interaction between two SSL branches is critical for capturing complementary representations and improving generalization. Replacing the original CNN–XLS-R attention with a bidirectional design also causes slight degradation on LA, DF, and ITW, indicating that the original design better models the complementary relationship between local signal-level features and high-level semantic representations.

\noindent \textbf{Impact of Different Adversarial Bona Fide Purification Hyperparameters}

\begin{table}[!t]
\centering
\caption{Effect of different bona fide application ratios under a fixed 2000 Hz cutoff frequency. Results are reported using EER (\%) and min t-DCF.}
\label{tab:saniboost_ratio}
\resizebox{0.6\linewidth}{!}{
\begin{tabular}{c|c|c|c}
\toprule
\textbf{Ratio} & \textbf{DF EER$\downarrow$} & \textbf{LA EER/min t-DCF$\downarrow$} & \textbf{ITW EER$\downarrow$} \\
\midrule
0.05 & 1.94 & 1.65 / 0.2314 & 7.96 \\
0.15 & 1.82 & 2.85 / 0.2550 & 6.47 \\
0.25 & 1.91 & 5.33 / 0.3299 & 6.51 \\
0.35 & \textbf{1.31} & 1.88 / 0.2359 & \textbf{5.44} \\
0.45 & 2.28 & 2.61 / 0.2573 & 5.63 \\
0.50 & 1.94 & 3.30 / 0.2719 & 6.97 \\
0.55 & 2.03 & 2.86 / 0.2642 & 6.84 \\
0.65 & 2.02 & 2.18 / 0.2459 & 7.29 \\
0.75 & 2.29 & \textbf{1.53} / \textbf{0.2271} & 8.30 \\
0.85 & 2.21 & 2.54 / 0.2487 & 10.49 \\
0.95 & 2.87 & 2.40 / 0.2482 & 9.73 \\
1.00 & 43.18 & 31.04 / 0.8075 & 6.05 \\
\bottomrule
\end{tabular}
}
\end{table}

\begin{table}[!t]
\centering
\caption{Effect of different high-pass filter cutoff frequencies under a fixed 0.35 application ratio. Results are reported using EER (\%) and min t-DCF.}
\label{tab:saniboost_cutoff}
\resizebox{0.9\linewidth}{!}{
\begin{tabular}{c|c|c|c}
\toprule
\textbf{Cutoff Frequency (Hz)} & \textbf{DF EER$\downarrow$} & \textbf{LA EER/min t-DCF$\downarrow$} & \textbf{ITW EER$\downarrow$} \\
\midrule
0    & 1.59 & 3.96 / 0.2936 & 6.42 \\
1000 & 1.92 & 2.74 / 0.2555 & 6.71 \\
2000 & \textbf{1.31} & \textbf{1.88} / \textbf{0.2359} & 5.44 \\
3000 & 2.01 & 3.55 / 0.2835 & 5.95 \\
4000 & 1.69 & 4.40 / 0.3025 & 5.42 \\
5000 & 1.87 & 4.09 / 0.2923 & \textbf{4.98} \\
6000 & 1.45 & 2.90 / 0.2663 & 6.24 \\
7000 & 1.63 & 2.74 / 0.2556 & 6.11 \\
\bottomrule
\end{tabular}
}
\end{table}

Bona Fide Purification acts as a controlled clean-speech regularization strategy avoiding the blurring of real or fake decision boundaries. 

Unlike conventional denoising approaches that may suppress high-frequency harmonics, this method explicitly preserves high-frequency information through a  3rd-order high-pass filter with a cutoff frequency of 2000 Hz. In addition, a lightweight noise gate (0.02$\times$ maximum amplitude) and soft compression are adopted to retain natural acoustic properties while removing low-frequency environmental shortcut cues.

To further analyze the effect of the hyperparameters, we conduct additional experiments by varying both the bona fide application ratio and the high-pass filter cutoff frequency. Notably, \textbf{these experiments were conducted only after the main experimental results had been finalized}, and the results of this analysis were not used for hyperparameter selection or tuning of SaniBoost. The results are summarized in Tables~\ref{tab:saniboost_ratio} and~\ref{tab:saniboost_cutoff}.
\noindent
From Table~\ref{tab:saniboost_ratio}, we observe that the proportion of Purification has a significant impact on model generalization. When the ratio is too small, the model still suffers from environmental shortcut bias, leading to relatively weaker robustness on ITW and DF. As the ratio increases, the performance gradually improves. However, when the ratio becomes excessively large (e.g., 0.75--1.0), the performance starts to degrade substantially. Especially at ratio = 1.0, the model collapses on LA, indicating that over-regularizing all bona fide samples severely damages the natural distribution and blurs the real/fake decision boundary.

\subsection{Additional Studies under Controlled Model Capacity}
\label{app:controlled_capacity}

\paragraph{Objective and controlled alternatives.}
To examine whether GLAD's performance gains can be explained by
pretrained-backbone diversity or parameter count alone, we compare
its fusion and aggregation architecture with five controlled
alternatives while retaining the same fully fine-tuned XLS-R 300M
and WavLM-Large encoders and the signal-level CNN branch:
(i) naive dual-SSL + CNN fusion without cross-attention;
(ii) static SLS with sample-independent layer weights followed by
late fusion;
(iii) generic cross-attention with squeeze-and-excitation (SE) gating;
(iv) a parameter-matched large MLP operating on the naive fusion
features; and
(v) MFA-style pooling followed by late fusion.
These controls test whether simpler feature combination,
conventional attention and gating, or additional classifier
capacity can match GLAD under a common training pipeline.

\paragraph{Common training and selection criteria.}
All models use the same ASVspoof 2019 LA training and development splits, input preprocessing, SaniBoost augmentation, weighted
classification loss, AdamW optimizer, and batch size of five. Hyperparameters are selected using the same development-set-based
selection protocol, while allowing model-specific selected values. For every model, the checkpoint with the lowest development EER
is retained, with development loss used to break ties. Early stopping uses a common patience of five epochs.
The selected hyperparameters are then fixed across three training runs with seeds 909, 1234, and 1235. No target-domain evaluation data are used for hyperparameter or checkpoint selection.

\begin{table*}[t]
\centering
\caption{
Capacity-controlled comparison of GLAD and five alternative fusion architectures.
Detection performance is reported as mean $\pm$ sample standard deviation of EER (\%)
over three random seeds (909, 1234, and 1235).
All variants use the same backbones. Computational cost is reported in terms of parameter count, peak training memory, and FLOPs.
}
\label{tab:efficiency_comparison}

\small
\setlength{\tabcolsep}{4.5pt}
\renewcommand{\arraystretch}{1.10}

\resizebox{\textwidth}{!}{%
\begin{tabular}{lcccccc}
\toprule

\textbf{Model}
& \shortstack{\textbf{Total / Trainable}\\
              \textbf{Parameters (M)}}
& \shortstack{\textbf{Peak Training}\\
              \textbf{Memory (GiB)}}
& \textbf{FLOPs (G)}
& \shortstack{\textbf{21LA}\\
              \textbf{EER (\%)} $\downarrow$}
& \shortstack{\textbf{21DF}\\
              \textbf{EER (\%)} $\downarrow$}
& \shortstack{\textbf{ITW}\\
              \textbf{EER (\%)} $\downarrow$} \\

\midrule

Naive dual-SSL + CNN fusion
& 633.89 / 633.89
& 15.58
& 375.01
& $6.50 \pm 0.38$
& $3.57 \pm 0.19$
& $7.05 \pm 0.91$ \\

Static SLS + late fusion
& 633.90 / 633.90
& 15.76
& 375.02
& $5.73 \pm 0.50$
& $3.23 \pm 0.54$
& $7.40 \pm 3.00$ \\

Generic cross-attention + SE
& 642.55 / 642.55
& 15.67
& 378.39
& $4.42 \pm 2.10$
& $12.40 \pm 12.19$
& $23.69 \pm 15.02$ \\

Parameter-matched large MLP
& 662.24 / 662.24
& 15.89
& 375.10
& $4.87 \pm 1.34$
& $2.74 \pm 0.73$
& $6.63 \pm 0.84$ \\

MFA-style pooling + late fusion
& 646.71 / 646.71
& 16.74
& 377.42
& $6.00 \pm 0.21$
& $3.79 \pm 0.87$
& $8.27 \pm 1.56$ \\

\midrule

\textbf{GLAD}
& 662.23 / 662.23
& 16.14
& 381.58
& $\mathbf{2.70 \pm 0.74}$
& $\mathbf{1.53 \pm 0.29}$
& $\mathbf{5.91 \pm 0.61}$ \\

\bottomrule
\end{tabular}%
}

\end{table*}

\paragraph{Detection performance.}
As shown in Table~\ref{tab:efficiency_comparison},
GLAD achieves the lowest mean EER on all three benchmarks:$2.70 \pm 0.74$\% on 21LA,$1.53 \pm 0.29$\% on 21DF, and $5.91 \pm 0.61$\% on ITW. Naive fusion, static layer weighting with late fusion, and MFA-style pooling do not match GLAD's mean performance.
Replacing the proposed architecture with generic CA and SE gating also yields higher mean EERs and substantial variation across runs, particularly on 21DF and ITW.

The parameter-matched large MLP provides a particularly direct capacity control: it contains 662.24M parameters, compared with
662.23M for GLAD, but achieves mean EERs of 4.87\%, 2.74\%, and 6.63\% on 21LA, 21DF, and ITW, respectively.
GLAD therefore reduces mean EER by 2.17, 1.21, and 0.72 percentage points relative to this control. These results support the contribution of GLAD's fusion and aggregation design beyond pretrained-backbone diversity or parameter count alone.

To examine whether GLAD's performance gains can be explained by parameter count alone, we compare its fusion and aggregation architecture with five controlled alternatives while retaining the same fully fine-tuned XLS-R 300M and WavLM-Large encoders and the signal-level CNN branch: (i) naive dual-SSL + CNN fusion without cross-attention; (ii) static SLS with sample-independent layer weights followed by late fusion; (iii) generic cross-attention with squeeze-and-excitation gating; (iv) a parameter-matched large MLP operating on the naive fusion features; and (v) MFA-style pooling followed by late fusion. 
All model hypeparameter selection and other training details are totally same with GLAD. Also, we report the same three seed report. As shown in table \ref{tab:efficiency_comparison}, GLAD achieves the lowest mean EER on all three benchmarks. Notablely, the larger MLP model results shows although parameter larger than GLAD,  the performance even worse across all datasets. This results further support the contribution of our whole model design beyond using the same encoders.

\newpage

\end{document}

%% file: math_commands.tex
\usepackage{amsmath,amsfonts,bm}

\def\eqref#1{equation~\ref{#1}}

\def\1{\bm{1}}

\DeclareMathAlphabet{\mathsfit}{\encodingdefault}{\sfdefault}{m}{sl}
\SetMathAlphabet{\mathsfit}{bold}{\encodingdefault}{\sfdefault}{bx}{n}



%% file: Sections/Introduction.tex
\section{Introduction}
\vspace{-6pt}

Recent advances in generative AI and neural speech synthesis enable highly realistic and efficient speech generation. Unlike traditional pipelines that require extensive real data, modern zero-shot text-to-speech and voice conversion models can clone voices with fidelity using only a few seconds of reference speech~\cite{wang2023neural,le2023voicebox,shen2023naturalspeech,li2023styletts,qin2023openvoice}. As a result, synthetic speech has become increasingly indistinguishable from bona fide recordings to the human ear, posing severe risks to personal privacy, financial security, and individual reputation~\cite{mai2023warning,chesney2019deep,westerlund2019emergence}. To mitigate these threats, Speech Deepfake Detection (SDD) has emerged as a pivotal countermeasure, aiming to discriminate between bona fide speech and synthesized spoofing attacks~\cite{yi2023audio,li2024audio}.
However, the continuous evolution of synthesis algorithms has led to an explosion in the heterogeneity and subtlety of spoofing artifacts. As a result, existing detectors often overfit to specific artifacts present in the training distribution, leading to brittle boundaries that struggle to generalize. Consequently, they suffer severe performance degradation when confronted with unseen algorithms or out-of-distribution (OOD) scenarios characterized by varying channels and environmental noises~\cite{muller2022does}.

Currently, Self-Supervised Learning (SSL) models dominate feature extraction but suffer from a critical granularity mismatch. While their Automatic Speech Recognition (ASR)-driven objectives excel at modeling global semantic consistency, they inherently suppress non-semantic nuances, thereby overlooking fine-grained signal-level artifacts essential for detecting high-fidelity deepfakes. 
Although recent approaches attempt to reintroduce spectral details~\cite{wang2025awaveformer} or to use static physical priors~\cite{yang2024robust}, their late-fusion mechanisms fail to effectively integrate local cues with global contexts. Consequently, these detectors lack the representation completeness required to learn generalized attack patterns, rendering them brittle in the face of advanced synthesis algorithms.

\begin{figure}[t!]
    \centering
    \includegraphics[width=0.95\linewidth]{figures/comparison_proof_1120.png}
    \vspace{-1.0em}
    \caption{Resolving the Fine-grained Forgery of SSL Detectors. Saliency map~\cite{simonyan2013deep} visualization on a partial spoof sample from~\cite{zhang2022partialspoof}, where gray shaded areas denote the ground-truth spoofed regions. (A-B) The input waveform and spectrogram reveal high-frequency inconsistencies in the spoofed region. (C) The SSL baseline (XLS-R~\cite{zhang2024audio}) overlooks the artifact region. (D) In contrast, our GLAD accurately pinpoints the local anomaly with a distinct saliency peak. This comparison highlights that our Global-Local design effectively compensates for the local insensitivity of pure SSL backbones.}
    \label{fig:blind_spot}  
\vspace{-2.0em}
\end{figure}

Beyond feature extraction, the lack of adaptive representation selection severely limits generalization to unseen domains. Static hard-pruning strategies~\cite{el2025comprehensive} reduce redundancy but inevitably sever the semantic context needed for robustness. Conversely, while Sensitive Layer Selection (SLS)~\cite{zhang2024audio} attempt to compute input-conditioned layer weight. However, we find this model exhibits broadly similar dominant layer preferences across the examined target domains. Because discriminative clues shift significantly across attacks and channels, this rigidity prevents the model from dynamically recalibrating its focus. Consequently, existing methods fail to disentangle intrinsic spoofing traces from environmental noise, resulting in severe out-of-distribution (OOD) performance degradation.

Distinct from prior literature that merely hypothesizes these limitations, we conduct a pioneering empirical and visual analysis. To the best of our knowledge, we are the first to explicitly validate how two critical architectural vulnerabilities, namely global-feature dominance and domain-driven layer importance shifts, systemically degrade out-of-domain generalization. Driven by these original insights, we propose the Global-Local Adaptive Detector (GLAD), a unified framework comprising the Hierarchical Global-Local (HGL) backbone, the Hierarchical Adaptive Gating (HAG) mechanism, and the SaniBoost augmentation strategy.

Firstly, to resolve the granularity gap of single-view SSL, GLAD employs the HGL backbone. This module synergizes linguistic semantics with acoustic context via Layer-wise Cross-Attention, while Multi-Granularity Fusion grounds these global features using local irregularities from a raw-waveform CNN, capturing semantic and signal-level anomalies. Secondly, to enhance generalization, we introduce the Hierarchical Adaptive Gating (HAG) mechanism. Unlike domain-level weighting schemes, HAG modulates information flow by performing sample-wise adaptive re-weighting across multiple SSL layers and heterogeneous backbones, suppressing domain-related and spurious factors while amplifying the most discriminative spoofing evidence. Finally, to mitigate shortcut learning, we introduce SaniBoost. Through bona fide noise sanitization and signal normalization, this strategy is designed to discourage reliance on environmental shortcuts and improve detection robustness.

In summary, the contributions of this work are fourfold:
\begin{itemize}
    \item We conduct an empirical and visual analysis to explicitly validate that the OOD failure of SSL-based models stems from global-feature dominance and domain-driven layer importance shifts. Driven by these original insights, we propose GLAD, a unified framework that bridges the granularity gap between abstract semantics and signal-level cues. By integrating global semantic coherence with fine-grained local artifacts and dynamically recalibrating their importance at the sample level, GLAD significantly enhances generalization against OOD scenarios.
    
    \item The Hierarchical Global-Local Backbone (HGL) unifies disjoint feature views. Through hierarchical integration of deep semantic representations and acoustic contexts with raw-waveform signal irregularities, ensuring the comprehensive capture of both semantic incoherence and spectral anomalies.
    
    \item Unlike static fusion, the Hierarchical Adaptive Gating (HAG) mechanism dynamically modulates features by performing sample-wise adaptive re-weighting over multiple layers and heterogeneous representations.
    
    \item SaniBoost is a sanitized augmentation strategy that enhances robustness against unseen acoustic variations introduced by diverse recording and processing conditions. Extensive experiments show that GLAD achieves state-of-the-art (SOTA) performance across five benchmarks.
   
\end{itemize}
\vspace{-16pt}
\label{sec: Introduction}

%% file: Sections/RelatedWorks.tex
\section{Related Works}
\begin{figure}[t!]
    \centering
    \includegraphics[width=0.95\linewidth]{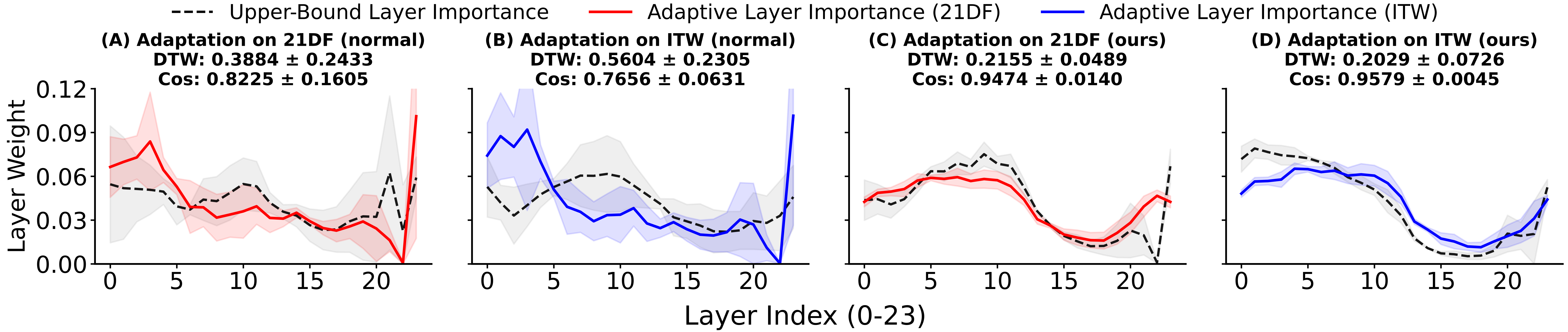}
    \caption{Visualization of layer weight distributions of the same SSL models (WavLM~\cite{chen2022wavlm}) using three different random seeds . Panels (A–B) display the weights with Sensitive Layer Selection (SLS)~\cite{zhang2024audio}, while Panels (C–D) are replaced by our proposed HAG. The solid lines (Red for 21DF, Blue for ITW) represent the weights assigned by the model only trained on the source domain (19LA), whereas the grey dashed lines indicate the oracle distributions derived from in-domain training (upper bound). The relatively high Dynamic Time Warping (DTW) scores in (A–B) reveal a significant misalignment between the static selection and the ideal layer preference. In contrast, the lower DTW scores in (C–D) demonstrate that HAG achieves a much tighter alignment with the oracle distribution, effectively adapting to cross-domain scenarios.
    }
    \label{fig:with_only_a}  
    \vspace{-1.0em}
\end{figure}
\label{sec:Related Works}

\vspace{-8pt}
\subsection{Speech Deepfake Detection Architectures} \vspace{-6pt}
Speech deepfake detection (SDD) has evolved from hand-crafted descriptors to data-driven deep representations. Early approaches primarily relied on prior knowledge from Digital Signal Processing (DSP), manually designing acoustic features such as MFCC~\cite{todisco2018integrated}, LFCC~\cite{sahidullah2015comparison}, and CQCC~\cite{todisco2016new}, often combined with GMM classifiers as seen in the ASVspoof 2019 baseline~\cite{todisco2019asvspoof}. The paradigm then shifted to end-to-end models. RawNet2~\cite{tak2021end} pioneered learning discriminative representations directly from raw waveforms, while AASIST~\cite{jung2022aasist} further integrated Graph Neural Networks (GNNs) to jointly model spectral-temporal cues. Recently, SSL models dominate due to their robust semantic modeling capabilities. For instance, Guo et al.~\cite{guo2024audio} leveraged WavLM to extract contextual features, designing a multi-fusion attention classifier. To further exploit hierarchical information, Zhang et al.~\cite{zhang2024audio} employed XLS-R as a backbone and proposed SLS to assign learnable weights to different Transformer layers. While some recent works attempt to combine SSL with handcrafted features via concatenation~\cite{wang2025awaveformer}, they typically lack deep interaction between modalities. Consequently, existing methods may still struggle to bridge the granularity gap between semantic and signal-level cues, and adaptive mechanisms like SLS despite computing input-conditioned layer weights, exhibit broadly similar dominant static layer preferences.

\vspace{-10pt}
\subsection{Domain Generalization and Data Augmentation}
\vspace{-6pt}
To improve robustness against domain shifts, data augmentation strategies have been extensively employed. Conventional approaches primarily focus on simulating external disturbances, such as time-frequency perturbations~\cite{chen2020generalization,cohen2022study} and codec/channel simulations~\cite{chen2021ur,caceres2021biometric}. SOTA methods like RawBoost~\cite{tak2022rawboost} further introduce complex noise injection at the raw waveform level to model channel variability. However, blindly augmenting data does not guarantee robustness. These strategies primarily simulate environmental noise without accounting for the entanglement between channel biases and spoofing traces. Recent studies highlight that this oversight risks shortcut learning~\cite{geirhos2020shortcut}, where models overfit to environmental factors (e.g., silence patterns, background noise) rather than learning intrinsic forgery evidence.

%% file: Sections/Methods.tex
\section{Methods}

\subsection{Motivation}



\begin{figure}[t!]
    \centering
    \includegraphics[width=\textwidth]{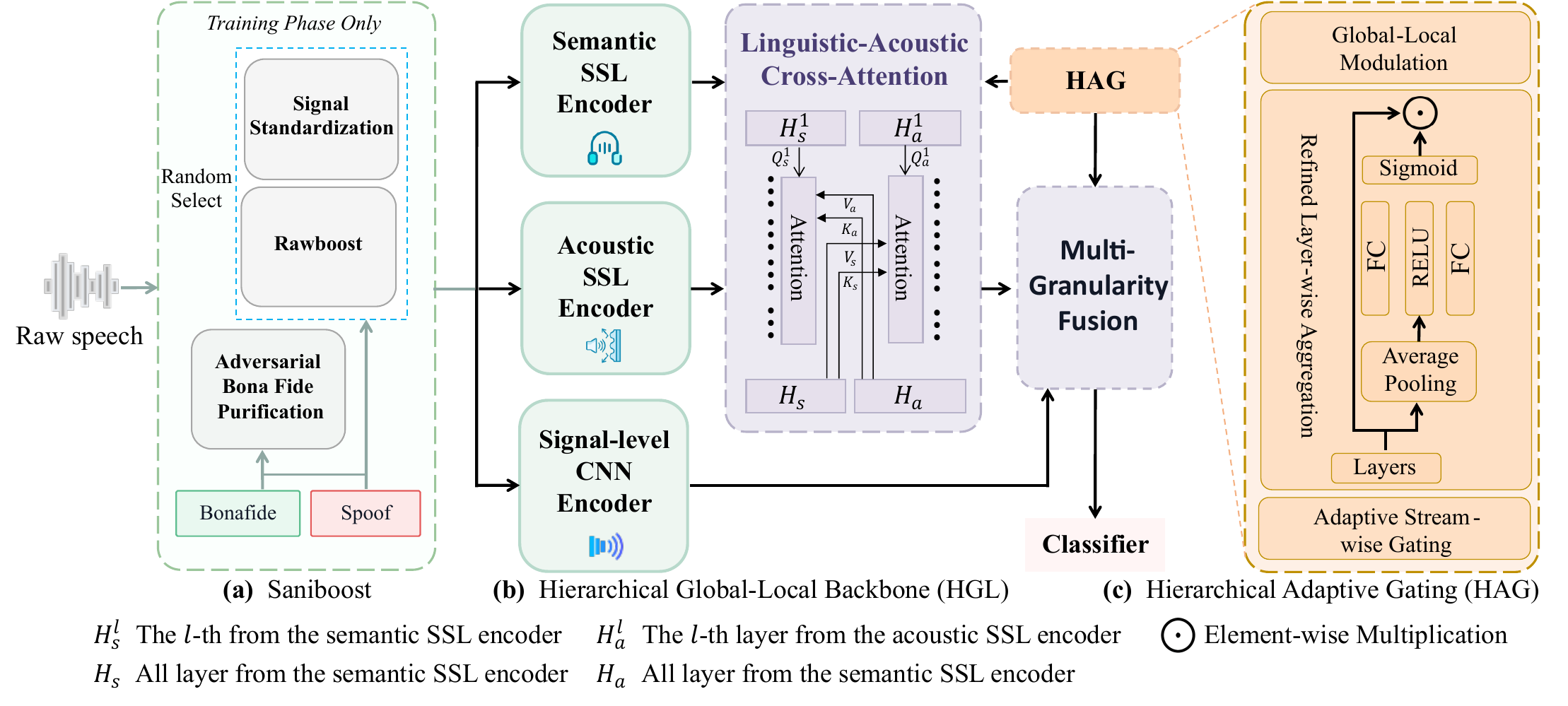}
    \caption{The overall architecture of the proposed GLAD framework from raw speech input to final classification. 
  \textbf{(a)} The Saniboost data augmentation method. 
  \textbf{(b)} The Hierarchical Global-Local Backbone (HGL) module. Here, $H_s$, $H_a$, and $\tilde{F}_c$ denote the representations extracted from the semantic SSL, acoustic SSL, and signal-level CNN encoders, respectively. $\hat{H}_s$ and $\hat{H}_a$ represent the Lingusitic-Acoustic Cross-Attention outputs where $H_s$ and $H_a$ serve as the Query, respectively. $F_{fused}$ refers to the linguistic-acoustic fused feature, $F_{refined}$ represents the refined fused feature, $F_g$ is the weighted refined fused feature, $F_u$ is the output of the Multi-Granularity Fusion, and $F_{final}$ denotes the final feature vector sent to the classifier.
  \textbf{(c)} The Hierarchical Adaptive Gating (HAG) module. $g_s$, $g_a$, $g_{sa}$, and $g_c$ represent the learned gating values for different streams. $\alpha_l$ denotes the learnable weights for layer-wise weighted summation.}
    \label{fig:fig1}
\end{figure}

\noindent \textbf{Capture Fine-grained Forgery.} Prior works indicate that SSL models tend to prioritize global semantic representations, often neglecting fine-grained acoustic details~\cite{qian2022contentvec}. However, discriminative cues in speech deepfakes are typically highly concentrated in local regions~\cite{sun2023aisynthesizedvoicedetectionusing}. To further investigate this limitation, we analyze representative SSL-based models~\cite{zhang2024audio} on a partial spoofing dataset~\cite{zhang2022partialspoof}, as illustrated in Figure~\ref{fig:blind_spot}. The empirical results reveal that conventional SSL-based methods exhibit limited sensitivity to localize these subtle local forgeries. Motivated by this observation, we propose the HGL backbone to extract features from diverse views.


\noindent\textbf{Adapt to Layer Importance Shift.} Recent studies indicate that discriminative cues in SSL models are non-uniformly distributed across network layers, with the layer relative importance varying across datasets~\cite{serrano2025improving,el2025comprehensive,martin2022vicomtech}. However, mainstream methods predominantly rely on heuristic selection or static aggregation weights. These static strategies tend to overfit the layer preference of the training set, thereby lacking the flexibility to adapt to the shifted layer importance in unseen domains.
To intuitively reveal this limitation, we conducted a preliminary investigation. We trained the SSL model with SLS mechanisms~\cite{zhang2024audio} on ASVspoof2019LA (19LA) and evaluated it on ASVspoof2021DF (21DF) and In-the-Wild (ITW). Additionally, to reveal the ideal optimal layer importance for each dataset, we trained the same models directly on 21DF and ITW, respectively. As illustrated in Figure~\ref{fig:with_only_a}, there is a significant discrepancy in layer importance across domains. Consequently, static strategies optimized for the source domain fail to generalize to unseen domains.
Driven by these findings, we propose the HAG mechanism to adaptively identify and emphasize the most informative layers and features in a sample-wise manner.


\subsection{Formulation of GLAD} 

As illustrated in Figure~\ref{fig:fig1}, GLAD accepts an input raw speech, which is processed by the Hierarchical Global-Local Backbone (HGL). To bridge the granularity gap, HGL synergizes multi-granular features through a hierarchical pipeline: it first employs Linguistic-Acoustic Cross-Attention to align deep linguistic semantics with acoustic contexts, followed by Multi-Granularity Fusion to ground these global representations with fine-grained local signal irregularities. Throughout this process, the Hierarchical Adaptive Gating (HAG) mechanism dynamically modulates feature contributions at both layers and branch levels. Finally, the unified representations are aggregated via ASP~\cite{okabe2018attentive} pooling, followed by a linear projection and a MLP classifier. Note that SaniBoost is only applied during the training phase.

\subsection{Hierarchical Global-Local Backbone}
To comprehensively characterize deepfake traces, HGL operates on a tri-branch input of three complementary feature views: deep linguistic semantics $\mathbf{X}_{s}$ from the Semantic SSL Encoder, high-level acoustic context $\mathbf{X}_{a}$ from the Acoustic SSL Encoder, and fine-grained signal irregularities $\mathbf{F}_{c}$ from the Signal-level CNN Encoder. Where $\mathbf{X}_{s}, \mathbf{X}_{a} , \mathbf{F}_{c} \in \mathbb{R}^{T \times D}$.
Unlike previous methods that independently process or simply concatenate these features, HGL bridges the granularity gap through a hierarchical interaction mechanism. As shown in Figure~\ref{fig:fig1}, it first establishes a global consensus by synergizing the two SSL streams via Linguistic-Acoustic Cross-Attention, and subsequently grounds these abstract representations in signal-level reality via Multi-Granularity Fusion.

\noindent \textbf{Linguistic-Acoustic Cross-Attention.}
\label{sec:LACA}
Self-supervised models, while powerful, often focus on disjoint aspects of speech and linguistic content versus acoustic environment. To synthesize a comprehensive global representation, we introduce a layer-wise interaction mechanism between the Semantic and Acoustic SSL Encoders. 

To obtain compact layer-wise descriptors, we first apply temporal average pooling over the time dimension to the frame-level features extracted from each SSL layer, yielding pooled representations $\mathbf{H}_{s}^{(l)}, \mathbf{H}_{a}^{(l)} \in \mathbb{R}^{D}$ at the $l$-th layers of the Semantic and Acoustic SSL Encoders, respectively. We then stack the pooled representations of all $L$ layers as $\mathbf{H}_{s} = [\mathbf{H}_{s}^{(1)}, \dots, \mathbf{H}_{s}^{(L)}]$ and $\mathbf{H}_{a} = [\mathbf{H}_{a}^{(1)}, \dots, \mathbf{H}_{a}^{(L)}]$, where $\mathbf{H}_{s}, \mathbf{H}_{a} \in \mathbb{R}^{L \times D}$.

We treat these two views as mutual references to rectify distinct semantic inconsistencies. Specifically, we employ a bidirectional Cross-Attention (CA) mechanism where each stream queries the other to retrieve complementary cues.  $\mathrm{CA}_{\theta}$ denotes a cross-attention block with query residual connections, layer normalization, and a feed-forward network. The two directions share the same CA block parameters $\theta_L$:
\begin{equation}
\widehat{H}_s
=
\operatorname{CA}_{\theta_L}(H_s,H_a,H_a),
\qquad
\widehat{H}_a
=
\operatorname{CA}_{\theta_L}(H_a,H_s,H_s).
\end{equation}

Here, $\hat{\mathbf{H}}_{s}^{(l)}$ incorporates acoustic context into linguistic features, identifying potential prosodic mismatches, while $\hat{\mathbf{H}}_{a}^{(l)}$ enriches acoustic features with semantic dependencies.
These refined layer-wise features are then aggregated via the HAG mechanism (detailed in Sec.~\ref{sec:hag}) producing a unified global representation $\mathbf{F}_{g} \in \mathbb{R}^{T \times D}$, which serves as the semantic anchor for the subsequent stage.

\noindent \textbf{Multi-Granularity Fusion.} 
While $\mathbf{F}_{g}$ captures high-level traces, it inherently lacks the resolution to detect fine-grained signal-level artifacts. To address this, we perform Multi-Granularity Fusion to align and integrate the local signal features $\mathbf{F}_{c} \in \mathbb{R}^{T' \times D'}$ extracted by the Signal-level CNN Encoder. First, addressing the temporal mismatch ($T \neq T'$), we align the local features $\mathbf{F}_{c}$ to the global anchor $\mathbf{F}_{g}$ via linear interpolation, yielding the aligned local features $\tilde{\mathbf{F}}_{c} \in \mathbb{R}^{T \times D}$. Subsequently, to prevent the strong semantic signals from overshadowing subtle signal-level anomalies, we employ a semantic-guided injection strategy. We use the global representation $\mathbf{F}_{g}$ as the \textit{Query} to interrogate the aligned local features $\tilde{\mathbf{F}}_{c}$ (serving as \textit{Key} and \textit{Value}) allowing the model to pinpoint signal irregularities that are inconsistent with the global semantic context:
\begin{equation}\mathbf{F}_{u} =  \text{CA}(\mathbf{Q}=\mathbf{F}_{g}, \mathbf{K}=\tilde{\mathbf{F}}_{c}, \mathbf{V}=\tilde{\mathbf{F}}_{c}).
\label{eq:MGF}
\end{equation}
By explicitly global semantics with local signal traces, this fused representation captures multi-scale spoofing evidence, ranging from high-level semantic incoherence to low-level waveform jitters.
\subsection{Hierarchical Adaptive Gating}
\label{sec:hag}
To enable adaptive detection across diverse deepfake attacks, we propose HAG to explicitly regulate feature sources, hierarchical contributions, and fusion ratios. HAG operates along three dimensions: stream-wise gating, layer-wise aggregation, and global-local modulation.

\noindent \textbf{Adaptive Stream-wise Gating.}
First, to synthesize the information from two SSL encoders at each layer $l$, we employ a gated residual mechanism. This allows the model to dynamically balance the contributions between the original features ($\mathbf{H}_{s}^{(l)}, \mathbf{H}_{a}^{(l)}$) and their interaction-refined counterparts ($\hat{\mathbf{H}}_{s}^{(l)}$ derived in Sec.~\ref{sec:LACA}).
We define sample-wise gates to weigh the importance of each stream:
\begin{equation}
\begin{aligned}
g_s^{(l)}
&=
\sigma\left(w_s^\top H_s^{(l)}+b_s\right),\\
g_a^{(l)}
&=
\sigma\left(w_a^\top H_a^{(l)}+b_a\right),\\
g_{sa}^{(l)}
&=
\sigma\left(w_{sa}^\top\widehat{H}_s^{(l)}+b_{sa}\right).
\end{aligned}
\end{equation}

where $\sigma$ denotes the sigmoid activation and $\mathcal{W}(\cdot)$ represents a lightweight linear projection. Since the final representation is anchored to the semantic backbone, acoustic cues may be attenuated when injected into semantic feature space. So to regulate the contribution of the acoustic branch, we introduce an additional gating mechanism on  $\hat{\mathbf{H}}_{s}^{(l)}$, allowing model to adaptively modulate the degree of acoustic information injection via CA.

The fused layer-level representation $\hat{\mathbf{F}}_{fused}^{(l)}$ is computed by combining the interaction terms and the gated original residual:

\begin{equation}
\mathbf{F}_{fused}^{(l)}
=
\hat{\mathbf{H}}_{s}^{(l)}
+
\hat{\mathbf{H}}_{a}^{(l)}
+
g_{s}^{(l)} \odot \mathbf{H}_{s}^{(l)}
+
g_{a}^{(l)} \odot \mathbf{H}_{a}^{(l)}
+
g_{sa}^{(l)} \odot \hat{\mathbf{H}}_{s}^{(l)}.
\end{equation}

Finally, to preserve the semantic structure while incorporating complementary cues,
We refined the fused representations within the semantic encoder,
yielding the refined fused feature $\mathbf{F}_{refined}^{(l)}$:
\begin{equation}
\begin{aligned}
\mathbf{F}_{refined}^{(l)}=\mathbf{F}_{fused}^{(l)}+\mathbf{X}_{s}^{(l)}  .\\
\end{aligned}
\end{equation}

This formulation ensures that the model can selectively retain original semantic/acoustic cues while integrating the cross-modal interaction gains.

\noindent \textbf{Refined Layer-wise Aggregation.}
After obtaining the refined-fused features $\mathbf{F}_{refined}^{(l)}$ for all $L$ layers, we perform weighted aggregation to distill the most discriminative hierarchical information. We refine the standard SLS strategy by replacing the single linear scoring layer with a Multi-Layer Perceptron (MLP) module. This design captures complex non-linear dependencies across different hidden states.
Specifically, the importance coefficient $\alpha_l$ for layer $l$ is computed via two linear transformations interconnected by a ReLU activation, followed by a sigmoid function, and the resulting weights are then used to aggregate the refined frame-level features:
\begin{equation}
\alpha_l
=
\sigma\left(
W_2\operatorname{ReLU}
\left(W_1F_{\mathrm{fused}}^{(l)}+b_1\right)
+b_2
\right).
\end{equation}

\noindent \textbf{Global-Local Modulation.}
To regulate the contribution of the local branch, we introduce a gate mechanism at the final fusion stage. Recall that $\mathbf{F}_{u}$ is the output of the Multi-Granularity Fusion (Eq.~\ref{eq:MGF}). We utilize the aligned signal features $\tilde{\mathbf{F}}_{c}$ to compute a modulation gate $g_{c}$:
\begin{equation}
g_{c} = \sigma(\mathcal{W}(\tilde{\mathbf{F}}_{c})).
\end{equation}
The final representation $\mathbf{F}_{final}$ is obtained by injecting gated local signal cues:

\begin{equation}
\mathbf{F}_{final} = \mathbf{F}_{u} + g_{c} \odot \tilde{\mathbf{F}}_{c}.
\end{equation}

This allows GLAD to adaptively amplify fine-grained artifacts when semantic traces are ambiguous.

\subsection{Saniboost}

Existing deepfake detectors are prone to shortcut learning, often overfitting to environmental nuances (e.g., loudness, silence patterns, or channel noise) rather than intrinsic forgery traces. To mitigate these spurious correlations, we introduce SaniBoost, a composite data augmentation strategy applied exclusively during the training phase.

\paragraph{Signal Standardization.} To eliminate physical biases, we first apply rigorous signal normalization.
\textbf{Loudness Invariance:} We implement Root Mean Square (RMS) normalization to standardize the energy levels. This encourages the model to discriminate based on spectral content rather than absolute signal amplitude.
\textbf{Phase Robustness:} We apply random Phase Inversion ($x \leftarrow -x$) and inject perturbation Gaussian noise. This prevents the detector from memorizing specific waveform polarities or pristine digital signatures, enhancing robustness against trivial signal variations.

\paragraph{Adversarial Bona Fide Purification.}  

Standard datasets may exhibit recording-condition biases, where bona fide speech contains more background noise than synthesized audio. Models may exploit this correlation by treating background noise as a feature of authenticity. To discourage reliance on this potential shortcut, we propose a counterintuitive Hyper-Clean Strategy. We apply aggressive high-pass filtering and noise gating to a subset of \textbf{bona fide} samples while retaining their original labels. By attenuating low-frequency energy and low-amplitude components, we artificially create ``hard positive'' samples that must still be recognized as bona fide despite these substantial acoustic changes. This augmentation is intended to discourage the HGL backbone from over-relying on environmental context and improve detection robustness under varying recording conditions.


%% file: Sections/Experiments.tex
\vspace{-10pt}
\section{Experiments}
\vspace{-10pt}
\subsection{Experiment Setups}
\vspace{-8pt}

\noindent\textbf{Datasets.} To evaluate the generalization capability of our proposed method against unseen attacks and OOD scenarios, we trained our model on the ASVspoof 2019 LA training set and evaluated it on the official evaluation partitions of ASVspoof 2019 LA (19LA), ASVspoof 2021 LA (21LA), ASVspoof 2021 DF (21DF), Fake-or-Real (FoR), and PartialSpoof. For In-The-Wild (ITW), we evaluated on the full dataset.

\begin{table*}[t]
\centering
\caption{Performance comparison across multiple benchmarks, including ASVspoof 2021 LA/DF, PartialSpoof, and real-world scenarios (ITW and FoR). Lower values are better ($\downarrow$).}
\label{tab:overall_results}
\small
\resizebox{\textwidth}{!}{%
\begin{tabular}{lccccccl}
\toprule
\multirow{2}{*}{\textbf{Method}} 
& \multicolumn{2}{c}{\textbf{2021 LA}} 
& \textbf{2021 DF} 
& \textbf{PartialSpoof} 
& \textbf{ITW} 
& \textbf{FoR} \\
\cmidrule(lr){2-3}
& \textbf{EER (\%)$\downarrow$} 
& \textbf{min t-DCF$\downarrow$} 
& \textbf{EER (\%)$\downarrow$} 
& \textbf{EER (\%)$\downarrow$} 
& \textbf{EER (\%)$\downarrow$} 
& \textbf{EER (\%)$\downarrow$} \\
\midrule
AASIST           & 12.82 & 0.5468 & 20.25 & 33.02  & 43.50 & 44.26 \\
W2V2-ST          & 7.36  & 0.3769 & 8.13  & 10.96 & 18.67 & 8.98 \\
AMSDF            & 2.00  & 0.2408 & 3.82  & 14.07 & 15.45 & 11.44 \\
XLS-53 \& LGF    & 6.53  & 0.3400 & 4.75  & 9.25 & 26.37 & 21.90 \\
XLS-R \& SLS     & 4.47  & 0.2907 & 2.30  & 8.28 & 7.38  & 5.15 \\
XLS-R\& MultiConv  & 6.01  & 0.3383 & 6.66 & 13.26 & 7.38  &  27.19 \\
STCA\& LMDC      & 2.05  & 0.2416 & 6.36 & 14.80 & 20.35  &  16.54 \\
XLSR-Conformer \& TCM   &3.60  & 0.2846 & 2.75 & 7.38 & 10.54  & 40.62 \\
WaveSpec  &  15.32  & 0.4844 & 16.66 & 28.94 & 23.75  & 41.73 \\
All-typed Add  &  10.51 & 0.4546 & 6.04 & 8.85 & 9.66  & 5.51 \\
WaveSP  &  6.81 & 0.3588 & 3.95 & 9.31 & 8.67  & 11.95 \\
\midrule

\textbf{GLAD (Ours)} 
& \textbf{1.88} 
& \textbf{0.2359} 
& \textbf{1.31} 
& \textbf{6.53} 
& \textbf{5.44} 
& \textbf{1.10} \\

\bottomrule
\end{tabular}%
}
\vspace{-0.5em}
\end{table*}

\noindent \textbf{Evaluation Metrics and Benchmarks.}~Following standard protocols, we report performance via Equal Error Rate (EER) and minimum tandem Detection Cost Function (min t-DCF).
For a comprehensive comparison, we evaluate our method against numerous representative spoofing detection approaches, broadly categorized into three groups based on model design and training paradigms:

\noindent \textit{Handcrafted-based Methods.}~This category includes classical approaches built upon manually designed acoustic features and conventional back-ends, such as LC-Res18~\cite{ma2023end}.

\noindent \textit{End-to-End Deep Learning Models.}~We also compare against end-to-end neural models (convolutional, graph-based, or multi-view) trained from scratch, including: TSSDN~\cite{hua2021towards}, RawGAT-ST~\cite{tak2021end}, AASIST~\cite{jung2022aasist}, and AMSDF~\cite{wu2024audio}.

\noindent \textit{SSL-based Methods.}~Finally, we consider recent approaches that exploit large-scale SSL models as front-ends, including W2V2-ST~\cite{tak2022automatic}, XLS-R \& SLS~\cite{zhang2024audio}, XLS-R\& MultiConv~\cite{tran2025multi}, XLS-53 \& LGF~\cite{wang2021investigating}, STCA\& LMDC~\cite{hao2025integrating}, XLSR-Conformer \& TCM~\cite{truong2024temporal}, and WaveSpec~\cite{jin2025wave},All-typed Add~\cite{xie2026detect} and WaveSP~\cite{xuan2026wavesp}.

\noindent \textbf{Implementation Details.} All experiments are conducted on a single NVIDIA RTX 6000 Ada GPU. We adopt XLS-R 300M~\cite{babu2021xls} as the semantic encoder and WavLM-Large~\cite{chen2022wavlm} as the acoustic encoder and a signal-level CNN encoder~\cite{ravanelli2018speaker}. Both SSL backbones are fully fine-tuned during training. We used the AdamW optimizer~\cite{kingma2014adam} with a weight decay of $1 \times 10^{-4}$. The learning rate was set to $1 \times 10^{-6}$ for the SSL backbones and $5 \times 10^{-6}$ for others. The batch size was set to 5, and the proposed SaniBoost strategy was employed for data augmentation. To mitigate class imbalance, we utilized a Weighted CE Loss with weights of $0.9$ for bona fide and $0.1$ for spoofed samples. Early stopping was applied with a patience of 5 epochs based on the 19LA development set.
All hyperparameters in this work were selected exclusively using the 19LA development set. To reduce computational cost during preliminary hyperparameter tuning, we randomly sampled 3,000 utterances from the 19LA training set and 3,000 utterances from the development set using a fixed random seed of 1234. This pilot subset corresponds to approximately one tenth of the full 19LA training data. Each candidate configuration was trained for five epochs, and the hyperparameters were selected according to the lowest development set EER.

For SaniBoost, we first evaluated bona fide application ratios of 0.15, 0.35, 0.55, and 0.75, and selected 0.35 based on the development set performance. We then fixed the application ratio at 0.35 and compared high-pass cutoff frequencies of 2,000, 4,000, and 6,000 Hz. A cutoff frequency of 2,000 Hz was selected. After hyperparameter selection, these settings were fixed for all subsequent experiments and evaluations.

\vspace{-20pt}
\subsection{Comparative Studies}
\vspace{-10pt}

\noindent \textbf{Robustness on 21LA \& 21DF.} Under strict OOD evaluation (trained only on 19LA), GLAD achieves SOTA results on both 21LA and 21DF sets (Table~\ref{tab:overall_results}). It reaches a remarkable \textbf{1.31\%} EER on the DF track, highlighting HGL's resilience to unknown codecs. Furthermore, GLAD outperforms single-stream baselines (e.g., XLS-R\& MultiConv), confirming the advantage of synergizing pre-trained linguistic-acoustic priors with fine-grained signal-level cues.


\noindent \textbf{Generalization to Real-world Scenarios.}
We further assess practical utility on the non-laboratory ITW and FoR datasets. Despite significant domain shifts, GLAD achieves EERs of \textbf{5.44\%} and \textbf{1.10\%}, respectively (Table~\ref{tab:overall_results}). This performance not only surpasses strong SSL baselines (e.g., XLS-R \& SLS at 7.38\% and 5.15\%) but also avoids the performance collapse seen in non-SSL models like AASIST (EER $> 30\%$). We attribute this robustness to GLAD's holistic design, which ensures representation completeness via multi-granular feature integration, while effectively decoupling environmental shortcuts and dynamically recalibrating the hierarchy to capture forensic evidence.
\noindent \textbf{Generalization to Partial Spoof Scenarios.}
We further evaluate GLAD on partial spoof datasets to assess its robustness in detecting partially spoofed audio. As shown in Table~\ref{tab:overall_results}, GLAD significantly outperforms all baseline methods on the PartialSpoof benchmark, with an EER of 6.53\%. This demonstrates that our model effectively detects partial spoofing attacks.

\begin{figure}[!t]
    \centering
    \includegraphics[width=0.95\linewidth]{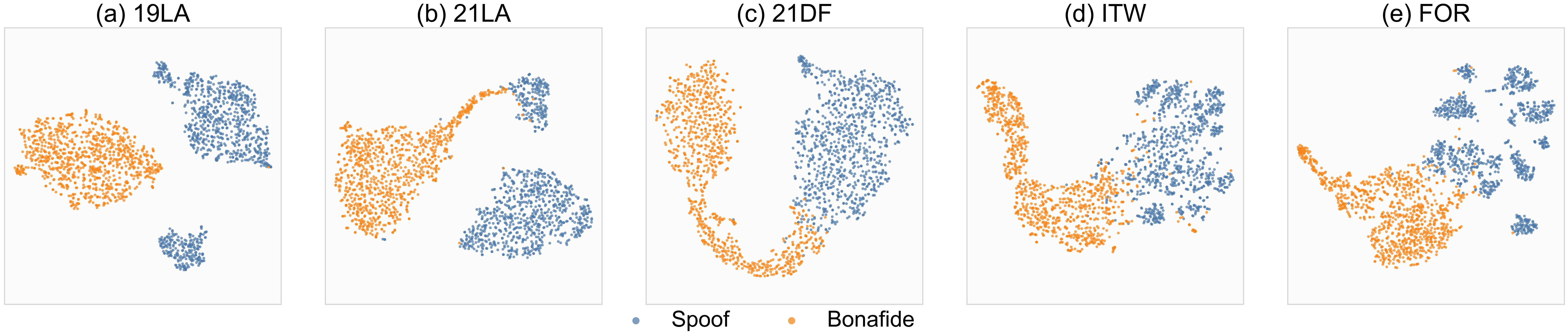}
    \caption{t-SNE visualization of feature embeddings obtained by GLAD. Orange denotes bona fide samples, and blue denotes spoofed samples (1000 samples per class). 
    }
    \label{fig:tsne_feat_visualization}  

\end{figure}

\begin{figure}[!t]
    \centering
    \includegraphics[width=0.95\linewidth]{figures/all_mechanisms_paper_ready_cut.png}
       \vspace{-0.9em}
    \caption{Analysis of learned gating distributions on different sets. The plot displays the distributions of component gates and the final aggregated importance coefficient (red solid line). The distinct activation patterns across different datasets confirm that the HAG module dynamically reconfigures its attention to capture domain-specific forensic traces while suppressing irrelevant signals.
    }
    \label{fig:prove_can_learn_different_feature_by_dataset}  
       \vspace{-0.9em}
\end{figure}

\noindent \textbf{Visualization of Feature Space.} To validate the discriminative power of the learned representations, we visualize the embedding space via t-SNE~\cite{JMLR:v9:vandermaaten08a}, showing in Figure~\ref{fig:tsne_feat_visualization}. Despite being trained solely on 19LA, GLAD maintains a stable geometric structure across all evaluation sets. The tight, clearly separated clusters for bona fide and spoofed samples corroborate that the model has successfully learned intrinsic forensic traces rather than overfitting to artifacts specific to the source domain.

\subsection{Ablation Study}
\begin{table}
\centering
\caption{Ablation study of main GLAD components on 21LA, 21DF, and ITW datasets (EER \%).}
\label{tab:ablation_modules}

\footnotesize
\setlength{\tabcolsep}{7pt}
\renewcommand{\arraystretch}{1.05}

\begin{tabular}{lccc}
\toprule
\textbf{Method} & \textbf{21LA$\downarrow$} & \textbf{21DF$\downarrow$} & \textbf{ITW$\downarrow$} \\
\midrule
w/o HGL       & 3.13 & 1.99 & 9.91 \\
w/o HAG       & 2.35 & 2.66 & 9.05 \\
w/o Saniboost & 4.62 & 2.47 & 6.44 \\
\midrule
\textbf{GLAD} & \textbf{1.88} & \textbf{1.31} & \textbf{5.44} \\
\bottomrule
\end{tabular}
\end{table}

GLAD is a rigorously designed framework in which each component is specifically intended to address a limitation of prior detectors. To rigorously validate the effectiveness and originality of this design, we performed comprehensive ablation studies in the 21LA, 21DF, and ITW datasets, demonstrating that each module is necessary and contributes uniquely. The results in Table~\ref{tab:ablation_modules} and Table~\ref{tab:ablation_submodules}.

\noindent \textbf{Effect of HGL.} Removing the HGL module and reverting to a single XLS-R backbone leads to substantial degradation (see row `w/o HGL' in Table~\ref{tab:ablation_modules}), particularly on the ITW set (EER: 5.44\% $\to$ 9.91\%). This confirms that relying on a single semantic view is insufficient; HGL's synergy of global linguistic-acoustic priors with local signal irregularities is essential for complex scenarios.

\noindent \textbf{Effect of HAG.} Removing the HAG module results in a significant EER increase across all test sets (see row `w/o HAG' in Table~\ref{tab:ablation_modules}), as the model loses the capability for dynamic feature regulation. Visualization in Figure~\ref{fig:prove_can_learn_different_feature_by_dataset} confirms domain-aware adaptation: ITW activates penultimate layers while DF targets middle layers, proving HAG reconfigures focus based on domain characteristics.

\noindent \textbf{Effect of SaniBoost.} Removing data augmentation (DA) leads to a substantial performance degradation across all evaluation settings (see row `w/o SaniBoost'in Table~\ref{tab:ablation_modules}), especially on 21LA (EER: 1.88\% $\to$ 4.62\%). These results support the contribution of SaniBoost to detection performance under OOD settings.

Additional ablations for individual components and controlled comparisons under matched model capacity are provided in Appendix~F. We further present quantitative analyses of each module in Appendix~E, together with additional visualizations and quantitative studies of gating behavior and alignment under partial spoofing in Appendices~D and~C.

%% file: Sections/Conclusion.tex
\vspace{-12pt}
\section{Conclusion and Limitations}
\vspace{-7pt}
This paper presented an investigation into the generalization failures of SSL-based SDD models, explicitly revealing critical vulnerabilities in feature granularity and layer adaptability. To resolve these foundational issues, we introduced GLAD, a robust framework that achieves SOTA cross-domain generalization through the synergy of HGL, HAG, and SaniBoost. By dynamically integrating multi-granularity integration with Saniboost augmentation, GLAD effectively improves detection performance across the evaluated domains. Despite these significant advancements, detecting unprecedented spoofing algorithms under extreme real-world noise remains an open challenge, occasionally leading to false predictions. Future research will focus on developing inherently generalized acoustic representations to minimize these deployment risks and further improve detection reliability.
\label{sec:conclusion}